\documentclass[11pt,a4paper]{article}
\pdfoutput=1

\usepackage{array}
\newcolumntype{P}[1]{>{\centering\arraybackslash}p{#1}}
\usepackage{textcomp}
\usepackage{siunitx}
\usepackage{subcaption}

\RequirePackage{ifpdf} 

\usepackage{cite} 
\usepackage{amsmath} 
\usepackage{mathtools}

\usepackage{jheppub}
\usepackage{pstricks}
\usepackage[final]{pdfpages} 
\usepackage{ifpdf} 
\usepackage{slashed}
\usepackage{hyperref}

\usepackage{color} 
\usepackage{graphics}

\usepackage{etoolbox} 
\usepackage{fixmath}

\usepackage{amsfonts}

\usepackage{multirow}
\usepackage{epstopdf}

\usepackage{relsize}
\usepackage{float}
\usepackage{soul}

\usepackage{tikz}
\usetikzlibrary{positioning,arrows}
\usetikzlibrary{decorations.pathmorphing}
\usetikzlibrary{decorations.markings}
\usetikzlibrary{shapes.geometric}
\usepackage{endnotes}
\tikzset{
    vector/.style={decorate, decoration={snake}, draw},
    provector/.style={decorate, decoration={snake,amplitude=2.5pt}, draw},
    antivector/.style={decorate, decoration={snake,amplitude=-2.5pt}, draw},
    fermion/.style={draw=black,
      postaction={decorate},decoration={markings,mark=at position .55
        with {\arrow[draw=black]{>}}}}, 
    fermionbar/.style={draw=black, postaction={decorate},
                       decoration={markings,mark=at position .55 with {\arrow[draw=black]{<}}}},
    fermionnoarrow/.style={draw=black},
    gluon/.style={decorate, draw=black,decoration={coil,amplitude=4pt, segment length=6pt}},
    scalar/.style={dashed,draw=black,
      postaction={decorate},decoration={markings,mark=at position .55
        with {\arrow[draw=black]{>}}}}, 
    scalarbar/.style={dashed,draw=black,
      postaction={decorate},decoration={markings,mark=at position .55
        with {\arrow[draw=black]{<}}}}, 
    scalarnoarrow/.style={dashed,draw=black},
    electron/.style={draw=black,
      postaction={decorate},decoration={markings,mark=at position .55
        with {\arrow[draw=black]{>}}}}, 
    bigvector/.style={decorate, decoration={snake,amplitude=4pt}, draw},
}

\renewcommand{\thefootnote}{\arabic{footnote}}

\newcommand{\be}{\begin{equation}}
\newcommand{\bdm}{\begin{displaymath}}
\newcommand{\bea}{\begin{eqnarray}}
\newcommand{\beastar}{\begin{eqnarray*}}

\newcommand{\ds}{\ensuremath{\delta_s}}

\newcommand{\ee}{\end{equation}}

\newcommand{\e}{\ensuremath{\epsilon}}

\newcommand{\edm}{\end{displaymath}}
\newcommand{\eea}{\end{eqnarray}}
\newcommand{\eeastar}{\end{eqnarray*}}

\newcommand{\mur}{\ensuremath{\mu_{R}}}
\newcommand{\muf}{\ensuremath{\mu_{F}}}

\newcommand{\rarrow}{\ensuremath{\rightarrow}}

\newcommand{\s}{\ensuremath{\sigma}}

\title{Higgs pair production from bottom quark annihilation to NNLO in QCD}
\author{Ajjath A H$^{a}$, Pulak Banerjee$^{a}$, Amlan Chakraborty$^{a}$, Prasanna K.\
  Dhani$^{a,b}$,  Pooja Mukherjee$^{a}$,  Narayan Rana$^{c,d}$ and V. Ravindran$^{a}$} 

\affiliation{$^a$ The Institute of Mathematical Sciences, HBNI, Taramani,
  Chennai 600113, India \\ 
$^b$ INFN, Sezione di Firenze, I-50019 Sesto Fiorentino, Florence, Italy\\
$^c$ Deutsches Elektronen--Synchrotron, DESY, Platanenallee 6, D-15738 Zeuthen, Germany\\
$^d$ INFN, Sezione di Milano, Via Celoria 16, I-20133 Milano, Italy} 

\emailAdd{ajjathah@imsc.res.in}
\emailAdd{bpulak@imsc.res.in}
\emailAdd{amlanchak@imsc.res.in}
\emailAdd{prasannakumar.dhani@fi.infn.it}
\emailAdd{poojamukherjee@imsc.res.in} 
\emailAdd{narayan.rana@mi.infn.it} 
\emailAdd{ravindra@imsc.res.in}

\abstract{We present the first results on the two-loop massless QCD corrections to the four-point amplitude  $b+\overline{b} \rightarrow H+H$  in the five flavor scheme, treating bottom quarks as massless. This amplitude is sensitive to the trilinear Higgs boson coupling. Our two-loop result for this amplitude constitutes of purely virtual contributions to the next-to-next-to-leading order QCD predictions for the production of a pair of Higgs bosons at the Large Hadron Collider.  
Using these two loop amplitudes and exploiting the universality of the soft contributions 
in perturbative QCD, we obtain the NNLO QCD effects 
in the soft plus virtual approximation.  We find that the inclusion of higher order terms reduce
the uncertainties resulting from the unphysical renormalisation and factorisation scales.
}
\begin{document}

\preprint{\quad IMSc/2018/11/09, DESY 18--166, TIF-UNIMI-2018-9}

\allowdisplaybreaks[4]
\unitlength1cm
\maketitle
\flushbottom

\let\footnote=\endnote
\renewcommand*{\thefootnote}{\fnsymbol{footnote}}


\section{Introduction}
Ever since the discovery of the Standard Model (SM) Higgs boson \cite{Chatrchyan:2012xdj,Aad:2012tfa}, one of the main objectives of 
the Large Hadron Collider (LHC) physics program has been 
to understand its properties. This involves the measurements of the Higgs boson couplings to the SM fermions and gauge bosons, 
its mass $(m_h)$, its CP properties etc.  Among these, 
the Higgs boson self couplings such as the trilinear ($\lambda_{3}^{\rm{SM}}$) and 
quartic couplings ($\lambda_{4}^{\rm{SM}}$) take prominence, which in the SM, can be unambiguously obtained from the Higgs boson mass. 
The SM Higgs potential, after the electro-weak symmetry breaking (EWSB), is given by
\begin{equation} \label{eq:lag}
\mathcal{L} \supset -\frac{m_{h}^2}{2}\phi^2(x) - \lambda_{3}^{\rm{SM}}v\phi^3(x) - \lambda_{4}^{\rm{SM}}\phi^4(x),
\quad  \lambda_{3}^{\rm{SM}} = \frac{m_{h}^2}{2v^2}, \quad \lambda_{4}^{\rm{SM}} = \frac{m_{h}^2}{8v^2},
\end{equation}
where $\phi(x)$ denotes the Higgs field. $v \approx 246$ GeV is the vacuum expectation value ({\it vev}) of the Higgs field and 
is fixed by the Fermi constant $G_{F}$.
The Higgs boson mass $m_{h}$,  is found  experimentally to be approximately equal to $125$ GeV and hence,  
the SM values for $\lambda_{3}^{\rm{SM}}$ and $\lambda_{4}^{\rm{SM}}$ are $\sim 0.13$ and $\sim 0.03$, respectively.
However, presence of beyond the SM (BSM) physics scenarios can modify these couplings, which, in turn, suggests 
independent measurements of them.  Any deviation from the SM values from the experimental measurements, 
could provide crucial information
on the structure of the scalar potential and thus could constrain 
BSM physics scenarios \cite{Englert:2014uua}. Moreover, the measurement of $\lambda_{3}^{\rm{SM}}$ also provides a way to check that 
the EWSB follows from the simple Ginzburg-Landau $\phi^4$ potential.
The observable that can probe these couplings at the hadron colliders is the production of multiple 
Higgs bosons \cite{Binoth:2006ym}. 
More precisely, the production of a pair of Higgs bosons can probe 
$\lambda_{3}^{\rm{SM}}$ but it is difficult to measure  
due to the smallness of its production cross section  and the presence of a
large QCD background. However, the study for the high luminosity LHC indicate that the Higgs boson pair production due to 
gluon fusion can predict $\lambda_{3}^{\rm{SM}}$ 
with $\mathcal{O}(1)$ accuracy 
\cite{ATL-PHYS-PUB-2014-019,ATL-PHYS-PUB-2015-046,CMS-PAS-FTR-15-002,CMS-DP-2016-064}. 
%

A pair of Higgs bosons can be produced through several partonic channels, viz gluon fusion, 
vector boson fusion, associated production with a vector boson or a pair of heavy quarks. 
Among these channels, the gluon fusion channel is the most dominant one at the LHC.
Being a loop-induced channel, gluon fusion gives a minuscule production cross section.
Additionally, the large background of this channel makes its measurement experimentally challenging.
Hence unless contributions from BSM physics enhance the production cross section, a measurement of this channel
will require a considerable integrated luminosity. On the other hand, in such a scenario, 
the sub-dominant channels 
in the SM could possibly become interesting as they would receive substantial contributions from new physics.
One such channel is the production of a pair of Higgs bosons in bottom quark annihilation. In certain 
supersymmetric models, namely the Minimal Supersymmetric SM (MSSM) \cite{Nilles:1983ge}, 
the bottom quark Yukawa coupling is enhanced {\it w.r.t.} the top quark Yukawa coupling, in the large $\tan\beta$ region,
where $\tan \beta$ is the ratio of {\it vev}'s of up and down type Higgs fields in the Higgs sector of the MSSM. 
Hence precise predictions for this channel is of high importance.


The dominant channel for Higgs boson pair production {\it i.e.} the gluon fusion channel, is mediated by a top quark loop.  This was evaluated at leading order (LO) in perturbative QCD 
in \cite{Glover:1987nx,Eboli:1987dy,Plehn:1996wb} decades before. The next-to-leading order (NLO) contributions were obtained in \cite{Dawson:1998py} 
only in the infinite top mass limit,
\textit{i.e.}~the top quark loop is integrated out resulting in an effective Lagrangian \cite{Dawson:1998py,Djouadi:1999gv,Djouadi:1999rca,Muhlleitner:2000jj} of gluons and Higgs fields.
There are several NLO results \cite{Grigo:2013rya,Frederix:2014hta,Maltoni:2014eza,Degrassi:2016vss,Grober:2017uho,Bonciani:2018omm} 
considering finite top quark mass effects which finally led to the full 
NLO corrections with exact top quark mass dependence \cite{Borowka:2016ehy, Borowka:2016ypz}.
In all these works, it has been found that, with an inclusive $K$-factor close to 2, 
the QCD corrections at NLO level are as large as that observed for a single Higgs boson production.
Hence, the next-to-next-to-leading order (NNLO) corrections were computed in \cite{deFlorian:2013jea},
in the effective theory, followed by a soft plus virtual (SV) approximated 
NNLO cross section in \cite{deFlorian:2013uza}. 
Consecutively, the effect of leading top quark mass corrections also has been included in \cite{Grigo:2015dia}.
Finally a fully differential distribution has been obtained at the NNLO level in \cite{Li:2013flc,Maierhofer:2013sha}
and also threshold resummation at next-to-next-to-leading logarithmic (NNLL) level in \cite{Shao:2013bz,deFlorian:2015moa}.
In \cite{Grazzini:2018bsd}, a re-weighting technique has been used to properly account for finite top mass effects at NNLO level.
Recently the virtual contributions relevant for next-to-next-to-next-to-leading order (N$^3$LO) QCD have also been 
computed in \cite{Banerjee:2018lfq}, within the effective theory.

While a plethora of work has been performed to reach ultimate precision for the gluon channel, the sub-dominant channels have not received
much attention. Although, as mentioned earlier, in certain BSM physics scenarios, they become consequential. We are particularly interested 
in the bottom quark annihilation channel where the Higgs boson couples to bottom quarks through the Yukawa coupling (proportional to the mass of the bottom quark), 
and the bottom quark is massless otherwise \cite{Aivazis:1993pi,Collins:1998rz,Kramer:2000hn}.
For single Higgs boson production through this channel, various work is known up to NNLO 
\cite{Dicus:1988cx,Dicus:1998hs,Maltoni:2003pn,Olness:1987ep,Gunion:1986pe,Harlander:2003ai} 
and N$^3$LO \cite{Ahmed:2014pka,Gehrmann:2014vha,Ahmed:2014cha,Ahmed:2014era} level
in the variable flavor scheme (VFS) \cite{Buza:1996wv,Bierenbaum:2009mv,Ablinger:2017err,Blumlein:2018jfm}.
For the production of a pair of Higgs bosons, the NLO correction was first obtained in \cite{Dawson:2006dm}.
Later on, NLO corrections have been obtained for this channel considering several 
BSM scenarios \cite{Dawson:2007wh,HongSheng:2005uy,Liu:2004pv}.  For the latter, the bottom quark annihilation
process dominates over the gluon fusion even at LO level. In addition,  their NLO QCD corrections are not
only sizeable but also larger than the supersymmetric QCD corrections.
In order to stabilize the cross section with respect to higher order radiative corrections, 
NNLO corrections to this channel are desirable. In this paper, as a first step, 
we present full NNLO QCD corrections from certain class of diagrams to inclusive cross section for producing
pair of Higgs bosons at the LHC and
apply soft plus virtual approximation for the other class of diagrams.  We find that the 
latter is sub-dominant and hence this approximation is good enough for 
phenomenological studies at the LHC. 

There are two classes of diagrams (we call them Class-A and Class-B, see Sec.~\ref{classification}), that contribute at two loops. The vertex type of diagrams 
which belong to Class-A are already known up to three loops \cite{Gehrmann:2014vha}.  For the Class-B, the one-loop 
QCD corrections exist in the literature~\cite{Dawson:2006dm}.  Here we compute the two-loop QCD corrections. 
%
We have studied the structure of infrared (IR) singularities and found that they are in agreement with the predictions by Catani~\cite{Catani:1998bh}. 
The finite results expressed in terms of classical polylogarithms of weight up to 4 is used to study the numerical stability of the amplitudes over a wide range of allowed kinematical variables.
These amplitudes constitute important component of NNLO predictions for the observables related to
the production of a pair of Higgs bosons at the LHC.  In general,    
these amplitudes suitably combined with the universal soft gluon contributions from the real
emission diagrams can be used to obtain soft plus virtual contributions up to NNLO level.  
We follow this approach \cite{Ravindran:2006cg} for the class B diagrams, while for the
class A diagrams we can suitably use results that are already available for the single production 
up to NNLO level. 
 

The paper is organized as follows.  In Sec.~\ref{sec2}, we discuss the Lagrangian, kinematics and
the classes of diagrams that are relevant for our computation.  Sec.~\ref{method} contains details of the computation,
the ultraviolet (UV) renormalization and the structure of IR divergences. 
We devote Sec.~\ref{numerical} for the numerical evaluation of the amplitude over a wide kinematic region.  
In sec.~\ref{inclusive}, we present 
relevant analytic results that are required to compute inclusive cross section for producing pair of Higgs bosons 
using the amplitudes computed up to two-loop level.
In Sec.~\ref{pheno}, we study their numerical impact at the LHC.  Finally, we conclude in Sec.~\ref{conclusion}.

\section{Theory} \label{sec2}
At the LHC, the dominant channel for the production of a pair of Higgs bosons is the
gluon fusion.  In addition, there are several sub-leading channels that contribute to
the production.   We consider one of these channels, 
namely the production through the bottom quark annihilation process.   
Since the LO and the NLO~\cite{Dawson:2006dm} QCD effects have already been studied in the
literature, as a first step towards the computation of the
full NNLO QCD corrections, we evaluate two-loop     
virtual contributions to the production of a pair of Higgs bosons in this channel. 
Note that we further need to compute contributions from real emission sub-process{es} to obtain IR safe
observables at the NNLO level.  These pure virtual corrections contribute to both the inclusive as well as the differential 
observables.   
These results along with the process independent soft gluon contributions, can give us the
first result at the NNLO level in the threshold limit, {\it i.e.}, when the invariant mass of the pair of Higgs bosons
approaches the partonic center of mass energy.

We use the regularized version of the QCD Lagrangian throughout.  The regularization scheme that we 
use, is the dimensional regularization (DR), 
in which all the fields and couplings of the Lagrangian and the loop integrals that appear 
in the Feynman diagrams 
are analytically continued to $d = 4 + \epsilon$ space-time dimensions.
In addition, we perform traces of Dirac $\gamma$ matrices in $d$-dimensions.

\subsection{The Yukawa interaction}  
We begin by reviewing the theoretical framework for the production of a pair of Higgs bosons 
via bottom quark annihilation at hadron colliders. The interaction part of the Lagrangian that is
responsible for the production is given by,
\begin{equation}
\mathcal{L}= -\lambda_{b} \phi(x)\bar{\psi}_{b}(x)\psi_{b}(x)\,,
\end{equation}
where $\psi_{b}(x)$ is the bottom quark field. 
$\lambda_{b}$ is the Yukawa coupling which after the EWSB is found to be ${m_{b}}/{v}$, 
where $m_{b}$ is the bottom quark mass and $v$ the {\it vev} of the Higgs field. 
In the SM, the ratio of the top quark Yukawa coupling ($\lambda_{t}$) 
and the bottom quark Yukawa coupling ($\lambda_{b}$) is found to be approximately 35 \textit{i.e.}
$\lambda_{t}/\lambda_{b} \approx 35$.  In addition, the bottom quark flux in the proton-proton
collision is much smaller than the gluon flux.  Hence, the contribution from this channel is sub-dominant as compared 
to the gluon fusion channel.  However, in the MSSM \cite{Nilles:1983ge},  
this ratio depends on the value of $\tan \beta$ which can increase the contribution resulting from the 
bottom quark annihilation channel.  At LO, 
\begin{eqnarray}
\frac{\lambda_{t}^{\rm MSSM}}{\lambda_{b}^{\rm MSSM}}
 = f_{\phi}(\alpha) \frac{m_t}{m_b} \frac{1}{\tan \beta} \,,
\end{eqnarray}   
with
\begin{equation}
f_{\phi}(\alpha) = 
         \begin{cases} -\cot \alpha\,\,\, \text{for}\,\,\, \phi = h,\\ 
                      ~~\tan \alpha\,\,\,\text{for}\,\,\, \phi = H, \\ 
                      ~~\cot \beta \,\,\,\,\text{for}\,\,\, \phi = A, 
         \end{cases} 
\end{equation}
where $h$ is the SM like light Higgs boson, $H$ and $A$ are the heavy and the pseudoscalar Higgs bosons, 
respectively.  The parameter $\alpha$ is the angle between weak and mass eigenstates of the neutral Higgs bosons $h$ and $H$.  Since, the bottom quark mass is much smaller than the other energy scales that appear
at the partonic level, we set the former to zero except in the Yukawa coupling in perturbation theory~\cite{Aivazis:1993pi, Collins:1998rz, Kramer:2000hn}.  
In particular, the finite mass effects from the bottom quarks are found to be suppressed 
by the inverse power of mass of the Higgs boson.  
The number of active flavors is taken to be $n_f =5$ and we work in the Feynman gauge. 
\subsection{Kinematics} \label{Not}
We compute all the relevant one- and two-loop amplitudes in perturbative QCD 
that contribute to the annihilation of bottom quarks into 
a pair of Higgs bosons. The scattering process
is given by   
\begin{equation}
\label{reactions}
b(p_1) + \bar{b}(p_2)\rightarrow H(p_{3}) + H(p_4) \,,
\end{equation} 
where $p_1$, $p_2$ are the momenta of incoming bottom, anti-bottom quarks and 
$p_3$, $p_4$ are the momenta of the final state Higgs bosons. 
The associated Mandelstam variables are,
\begin{eqnarray}
s = (p_1 + p_2)^2, \hspace{1cm} t = (p_1 - p_3)^2, \hspace{1cm} u = (p_2 - p_3)^2,
\end{eqnarray}
which satisfy the relation $s + t + u = 2m_{h}^2$. For convenience, we use the dimensionless variables $x, y$ and $z$ defined \cite{Gehrmann:2013cxs} as follows
\begin{equation}
s = m_{h}^2\frac{(1+x)^2}{x}, \hspace{1cm} t = -m_{h}^2y, \hspace{1cm} u = -m_{h}^2z \, .
\end{equation}
The variables $x, y$ and $z$ satisfy
\begin{equation}
 \frac{(1+x)^2}{x} - y - z = 2 \,.
\end{equation}
The final result will be expressed in term of logarithms and classical polylogarithms
which are functions of these scaling variables. 

\subsection{General structure of the amplitude}
The external states for the process given in Eq.~(\ref{reactions}) involve two fermions and two scalars,
hence the most general structure of the amplitude can be parameterized as
\begin{align}
\label{Amp}
\mathcal{A}_{ij} &= \bar{v}(p_2) \Big(\mathcal{C}_{1} + \mathcal{C}_{2}~ \slashed{p}_3 \Big) u(p_1)\delta_{ij}
\nonumber\\
&\equiv \left(\mathcal{C}_{1} \mathcal{T}_1 + \mathcal{C}_{2} \mathcal{T}_2\right)\delta_{ij} \, ,
\end{align}
where the coefficients $\mathcal{C}_{m} \equiv \mathcal{C}_{m} (x,y,z)$ with $m=1,2$
are scalar functions. In color space, the amplitude is diagonal in the indices $(i, j)$ of the
incoming quarks. Since, we are interested in higher order QCD corrections, we have used
symmetries such as Lorentz covariance, parity and
time reversal invariances to parameterize the amplitude.
In addition we have dropped those terms that vanish when the bottom quarks are massless.
The coefficients $\mathcal{C}_{m}$,
$m=1,2$, 
can be determined from the amplitude $\mathcal{A}_{ij}$ by using appropriate projection operators 
denoted by $\mathcal{P}(\mathcal{C}_{m})$, {\it i.e.},
\begin{equation}
\label{C2}
\mathcal{C}_{m} = \frac{1}{N}\sum \mathcal{P} (\mathcal{C}_{m})\mathcal{A}_{ij}\delta_{ij} \, ,
\end{equation}
where the sum includes spin, flavors and colors of the external fermions; $N$ is the number of colors in SU(N) gauge theory.
In $d$-space-time dimensions, 
the projectors that satisfy $\sum\mathcal{P}(\mathcal{C}_m)\mathcal{T}_{m}=1$ and $\sum\mathcal{P}(\mathcal{C}_m)\mathcal{T}_{n}=0$
$\forall\, m\neq n$, are found to be
\begin{align} 
\label{projectors}
\mathcal{P}(\mathcal{C}_1) &= \frac{1}{2s} \mathcal{T}_1^{\dagger}\, ,
\nonumber\\ 
\mathcal{P}(\mathcal{C}_2) &= \frac{1}{2 [(m_h^2-t)(m_h^2 -u)-sm_h^2]}\mathcal{T}_2^{\dagger} \, . 
\end{align}
Since the application of projection operators on the amplitude gives only Lorentz scalar functions,
the algebraic manipulations with loop integrals become straightforward.  
The square of the amplitudes, that contributes to the total cross-section,
can now be obtained from the coefficients $\mathcal{C}_{1}$ and $\mathcal{C}_{2}$ using
\begin{align}
|\mathcal{A}_{ij}|^2 =N\left[ |\mathcal{C}_{1}|^2 \mathcal{T}_1\mathcal{T}_{1}^{\dagger} + |\mathcal{C}_{2}|^2 \mathcal{T}_2\mathcal{T}_2^{\dagger} +\mathcal{C}_{1}\mathcal{C}_{2}^\dagger \mathcal{T}_1\mathcal{T}_{2}^{\dagger} + \mathcal{C}_{1}^\dagger \mathcal{C}_{2}\mathcal{T}_2\mathcal{T}_{1}^{\dagger}\right] \, .
\end{align}
Note that these coefficients are in general complex due to the Feynman loop integrals. 
We expand the amplitude $\mathcal{A}_{ij}$ as well as the coefficients $\mathcal{C}_m$ in 
powers of the strong coupling constant defined by $a_s = g_s^2(\mu_R^2)/16 \pi^2$,  where $g_s$ is the renormalized strong
coupling constant and $\mu_R$ is the renormalization scale: 
\begin{equation}
 \mathcal{A}_{ij} = \sum_{l=0}^{\infty} a_s^l ~ \mathcal{A}^{(l)}_{ij} \,, 
 \quad
 \mathcal{C}_m = \sum_{l=0}^{\infty} a_s^l ~ \mathcal{C}^{(l)}_{m} \,,
\end{equation}
and consequently 
\begin{equation}
\label{C1C2}
\mathcal{A}^{(l)}_{ij} = \left(\mathcal{C}_{1}^{(l)} \mathcal{T}_{1} + \mathcal{C}_{2}^{(l)} \mathcal{T}_{2}\right)\delta_{ij} \,.
\end{equation}
Our next task is to compute these coefficients $\mathcal{C}_m^{(l)}$, $m=1,2$, up to two loop level, {\it i.e.}, up to $\mathcal{O}(a_s^2)$ in 
perturbative QCD.

\subsection{Classification of Feynman diagrams}
\label{classification}
At LO, only three Feynman diagrams contribute, 
out of which one contains single Yukawa and trilinear couplings, 
and the remaining ones are quadratic in the Yukawa coupling.
We denote the former by Class-A and the latter diagrams by Class-B.
The same classes of diagrams contribute beyond LO. 
 {We} elaborate on these classes of diagrams below:
\begin{itemize}

\item Class-A: It contains diagrams where an off-shell Higgs boson produced in the bottom quark annihilation
      process decays to a final state containing a pair of {on-shell} Higgs bosons ($H^*\rightarrow HH$) and 
      is proportional to $\lambda_3^{\rm SM} \lambda_b$. They are shown in Fig.~\ref{fig:classA}.
      Note that the decay part of the amplitudes does not get any QCD corrections, however the initial
      states do get.  These corrections
      are identical to those that contribute to the amplitudes for producing a single Higgs boson in bottom quark
      annihilation. The latter is known up to three-loop level in QCD \cite{Gehrmann:2014vha}.
\begin{figure}[h!]
\hspace{0.5cm}
\begin{tikzpicture}[line width=1 pt, scale=0.65]
\draw[fermion] (-1.2,1.2) -- (0,0);
\draw[fermion] (0,0) -- (-1.2,-1.2);
\draw[scalarnoarrow] (0,0) -- (1.5,0);
\draw[scalarnoarrow] (1.5,0) -- (2.5,1.0);
\draw[scalarnoarrow] (1.5,0) -- (2.5,-1.0);
\node at (-1.4,1.30) {$b$};
\node at (-1.4,-1.30) {$\bar{b}$};
\node at (0.85,0.35) {$H^{*}$};
\node at (2.8,1.05) {$H$};
\node at (2.8,-1.05) {$H$};
\hspace{5cm}
\draw[fermion] (-2.5,1.5) -- (0,0);
\draw[fermion] (0,0) -- (-2.5,-1.5);
\draw[gluon] (-1.9,-1.17) -- (-1.9,1.17);
\draw[scalarnoarrow] (0,0) -- (1.5,0);
\draw[scalarnoarrow] (1.5,0) -- (2.5,1.0);
\draw[scalarnoarrow] (1.5,0) -- (2.5,-1.0);
\node at (-2.7,1.6) {$b$};
\node at (-2.7,-1.6) {$\bar{b}$};
\node at (0.85,0.35) {$H^{*}$};
\node at (2.8,1.05) {$H$};
\node at (2.8,-1.05) {$H$};
\hspace{5cm}
\draw[fermion] (-2.5,1.5) -- (0,0);
\draw[fermion] (0,0) -- (-2.5,-1.5);
\draw[gluon] (-1.9,-1.17) -- (-1.9,1.17);
\draw[gluon] (-1.0,-0.63) -- (-1.0,0.63);
\draw[scalarnoarrow] (0,0) -- (1.5,0);
\draw[scalarnoarrow] (1.5,0) -- (2.5,1.0);
\draw[scalarnoarrow] (1.5,0) -- (2.5,-1.0);
\node at (-2.7,1.6) {$b$};
\node at (-2.7,-1.6) {$\bar{b}$};
\node at (0.85,0.35) {$H^{*}$};
\node at (2.8,1.05) {$H$};
\node at (2.8,-1.05) {$H$};
%
\end{tikzpicture}
\caption{Illustration of Class-A diagrams; Born, one and two-loop examples.}
\label{fig:classA}
\end{figure}
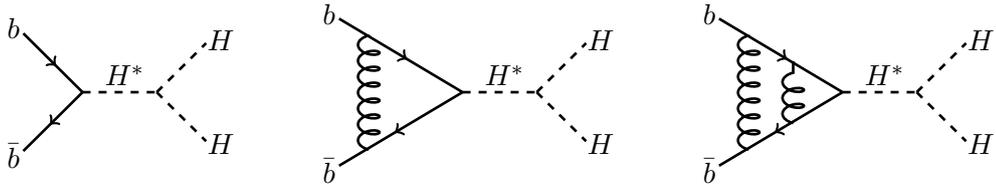
\item Class-B: 
      It contains diagrams where both the Higgs bosons couple directly to the bottom quarks  
      and hence they are  proportional to $\lambda_{b}^2$ as shown in Fig.~\ref{fig:classB}.  At two loops level,
      one encounters a new set of diagrams, the singlet contributions, where both the Higgs bosons are produced from a closed bottom quark loop
      as shown in Fig.~\ref{fig:classC}. Here in this class of diagrams, we have dropped the dominant contributions coming from the top quark loops and computed only those from bottom quark loops as the former ones are already included in the gluon initiated  subprocesses obtained in the heavy top limit in \cite{deFlorian:2013jea} for the Higgs pair production at the LHC.

\begin{figure}[h!]
\hspace{0.5cm}
\begin{tikzpicture}[line width=1 pt, scale=1]
\draw[fermion] (-2.0,1.0) -- (0,1);
\draw[fermion] (0,-1.0) -- (-2.0,-1);
\draw[fermion] (0,1) -- (0,-1);
\draw[scalarnoarrow] (0,1) -- (1.0,1);
\draw[scalarnoarrow] (0,-1) -- (1.0,-1);
\node at (-2.2,1.1) {$b$};
\node at (-2.2,-1.1) {$\bar{b}$};
\node at (1.2,1) {$H$};
\node at (1.2,-1) {$H$};
\hspace{5cm}
\draw[fermion] (-2.4,1.0) -- (-1.6,1);
\draw[fermion] (-1.6,1.0) -- (0,1);
\draw[fermion] (0,-1.0) -- (-1.6,-1);
\draw[fermion] (-1.6,-1.0) -- (-2.4,-1);
\draw[fermion] (0,1) -- (0,-1);
\draw[gluon] (-1.6,-1.0) -- (-1.6,1.0);
\draw[scalarnoarrow] (0,1) -- (1.0,1);
\draw[scalarnoarrow] (0,-1) -- (1.0,-1);
\node at (-2.6,1.1) {$b$};
\node at (-2.6,-1.1) {$\bar{b}$};
\node at (1.2,1) {$H$};
\node at (1.2,-1) {$H$};
\hspace{5cm}
\draw[fermion] (-2.4,1.0) -- (-1.6,1);
\draw[fermion] (-1.6,1.0) -- (-0.8,1);
\draw[fermion] (-0.8,1.0) -- (0,1);
\draw[fermion] (0,-1.0) -- (-0.8,-1);
\draw[fermion] (-0.8,-1.0) -- (-1.6,-1);
\draw[fermion] (-1.6,-1) -- (-2.4,-1);
\draw[fermion] (0,1) -- (0,-1);
\draw[gluon] (-1.6,-1.0) -- (-1.6,1.0);
\draw[gluon] (-0.8,-1.0) -- (-0.8,1.0);
\draw[scalarnoarrow] (0,1) -- (1.0,1);
\draw[scalarnoarrow] (0,-1) -- (1.0,-1);
\node at (-2.6,1.1) {$b$};
\node at (-2.6,-1.1) {$\bar{b}$};
\node at (1.2,1) {$H$};
\node at (1.2,-1) {$H$};
\end{tikzpicture}
\caption{Illustration of Class-B diagrams; Born, one and two-loop examples.}
\label{fig:classB}
\end{figure}
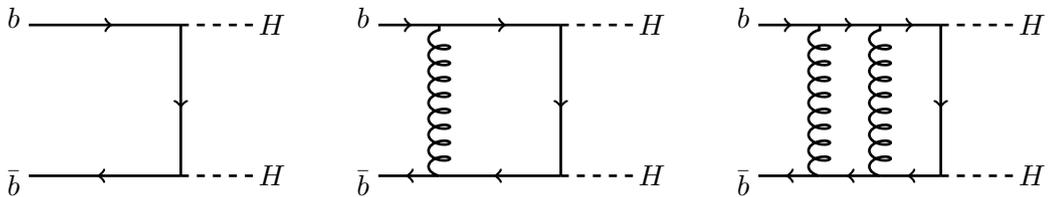

\begin{figure}[h!]
\begin{tikzpicture}[line width=1 pt, scale=1.0]
\hspace{2cm}
\draw[fermion] (-2.0,1.0) -- (-1.0,1);
\draw[fermion] (-1.0,-1.0) -- (-2.0,-1);
\draw[fermion] (-1.0,1) -- (-1.0,-1.0);
\draw[gluon] (-1.0,1.0) -- (0.5,1.0);
\draw[gluon] (-1.0,-1.0) -- (0.5,-1.0);
\draw[fermion] (0.5,1.0) -- (0.5,-1.0);
\draw[fermion] (0.5,-1.0) -- (1.5,-1.0);
\draw[fermion] (1.5,-1.0) -- (1.5,1.0);
\draw[fermion] (1.5,1.0) -- (0.5,1.0);
\draw[scalarnoarrow] (1.5,1) -- (2.0,1);
\draw[scalarnoarrow] (1.5,-1) -- (2.0,-1);
\node at (-2.2,1.1) {$b$};
\node at (-2.2,-1.1) {$\bar{b}$};
\node at (2.2,1) {$H$};
\node at (2.2,-1) {$H$};
\hspace{7cm} 
\draw[fermion] (-2.0,1.0) -- (-1.0,1);
\draw[fermion] (-1.0,-1.0) -- (-2.0,-1);
\draw[fermion] (-1.0,1) -- (-1.0,-1.0);
\draw[gluon] (-1.0,1.0) -- (1.0,1.0);
\draw[gluon] (-1.0,-1.0) -- (1.0,-1.0);
\draw[fermion] (1.0,1.0) -- (0.5,0.0);
\draw[fermion] (0.5,0.0) -- (1.0,-1.0);
\draw[fermion] (1.0,-1.0) -- (1.5,0.0);
\draw[fermion] (1.5,0.0) -- (1.0,1.0);
\draw[scalarnoarrow] (1.5,0.0) -- (2.0,0);
\draw[scalarnoarrow] (0.5,0) -- (0.0,0);
\node at (-2.2,1.1) {$b$};
\node at (-2.2,-1.1) {$\bar{b}$};
\node at (2.2,0) {$H$};
\node at (-0.2,0) {$H$};
%
%
%
\end{tikzpicture}
\caption{Illustration of special set of Class-B diagrams, the singlet contributions.} 
\label{fig:classC}
\end{figure}
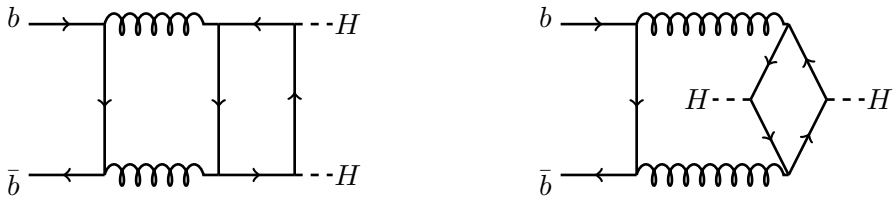
\end{itemize}
\section{Methodology}
\label{method}
\subsection{Computational details} 
\label{Calc}
It is easy to see from the form of $\mathcal{T}_i$ in Eq.~(\ref{Amp}), that only the Class-A diagrams contribute to $\mathcal{C}_1$ and 
the Class-B  to $\mathcal{C}_2$. Note that the Class-A diagrams are already computed to three loops in QCD \cite{Gehrmann:2014vha}. Hence in this section, we briefly discuss how the scalar function ${\mathcal{C}}_2$ in Eq.~(\ref{C2}) is 
computed order by order in perturbation theory. 
As we mentioned, we use the dimensional regularization, in which 
the space-time dimensions are taken to be $d=4 + \epsilon$ and perform 
traces of Dirac $\gamma$ matrices, contraction
of Lorentz indices in $d$-dimensions.
For convenience, we work with the bare form of the Lagrangian and evaluate
the coefficient ${\mathcal{C}}_2$ 
in powers of bare coupling constant $\hat a_s$, where $\hat{a}_s=\hat{ g}_s^2 /16 \pi^2$,  $\hat g_s$
being the dimensionless strong coupling constant.
Beyond LO, one- and two-loop amplitudes containing massless quarks, anti-quarks and gluons 
develop IR divergences in addition to UV ones.  There are two types of IR divergences, viz soft and collinear
divergences.   The soft ones are due to soft gluons and the collinear ones
arise due to massless quarks and gluons. 
Dimensional regularization regulates both these divergences in addition to UV divergences.

We have used QGRAF \cite{Nogueira:1991ex} to generate the Feynman diagrams at every order in the strong coupling constant.
Beyond one-loop, large number of Feynman diagrams contributes to the amplitude.
We find that there are 2 diagrams at the Born level, 10 diagrams at one-loop and 153 diagrams at two-loop level.
We then multiply these amplitudes with the projection operator 
$\mathcal{P}(\mathcal{C}_2)$
defined in Eq.~(\ref{projectors}) to obtain the scalar function ${\cal C}_2$.  
Substitution of Feynman rules and computation of various traces involving Dirac and Gell-Mann matrices, 
are done using in-house routines that use 
publicly available packages such as FORM \cite{Vermaseren:2000nd} and Mathematica.
At this stage we end up with a large number of  one- and two-loop Feynman integrals. 
The projection operators guarantee that all the tensor integrals
are converted to scalar integrals.   
We rearrange all the 
Feynman integrals into a few chosen integral families
through shifting of loop momentum. 
{ To achieve this, we use
the package Reduze2  \citep{vonManteuffel:2012np}.} At one-loop, the following three integral families can accommodate 
all the Feynman integrals
\begin{align} \label{basis1}
\left\lbrace\mathcal{P}_1, \mathcal{P}_{1:i}, \mathcal{P}_{1:i,i+1}, \mathcal{P}_{1:i,i+1,i+2}\right\rbrace \, ,
\end{align}
where, $i$ takes one of the values $\left\lbrace1,2,3\right\rbrace$ whose elements are arranged cyclically.
A typical two-loop topology contains at most seven propagators.  However, 
there are nine different Lorentz invariants $(k_i.k_j,k_i.p_j)$
which can appear in the numerator of an integral. Hence, we introduce two auxiliary propagators in each of the two-loop integral families.
The following two sets describe the six integral families that we use at two-loops, 
\begin{align} \label{basis2}
&\left\lbrace\mathcal{P}_0, \mathcal{P}_1, \mathcal{P}_2,\mathcal{P}_{1:i},\mathcal{P}_{2:i}, \mathcal{P}_{1:i,i+1},\mathcal{P}_{2:i,i+1} \mathcal{P}_{1:i,i+1,i+2}, \mathcal{P}_{2:i,i+1, i+2}\right\rbrace \, ,
\nonumber\\ 
&\left\lbrace\mathcal{P}_0, \mathcal{P}_1, \mathcal{P}_2,\mathcal{P}_{1:i},\mathcal{P}_{2:i}, \mathcal{P}_{1:i,i+1},\mathcal{P}_{2:i,i+1}, \mathcal{P}_{0:i+2}, \mathcal{P}_{1:i,i+1,i+2}\right\rbrace \, .
\end{align}
Here,
\begin{align*}
\mathcal{P}_\alpha = k_\alpha^2 \, , \ \ \  \mathcal{P}_{\alpha:i} = (k_\alpha- p_i)^2 \, , \ \ \ \mathcal{P}_{\alpha:ij} = (k_\alpha - p_i - p_j)^2 \, , \ \ \ \mathcal{P}_{\alpha:ijk} = (k_\alpha-p_i -p_j - p_k)^2 \, ,
\end{align*}
\begin{align*}
\mathcal{P}_0 = (k_1 - k_2)^2 \, , \ \ \ \ \  \mathcal{P}_{0:i} = (k_1 -k_2 -p_i)^2 \, .
\end{align*}
This large number of Feynman integrals belonging to different integral families can be
written in terms of a smaller set of integrals, so-called master integrals (MIs). 
This can be achieved by using the integration-by-parts (IBP) \cite{Tkachov:1981wb,Chetyrkin:1981qh} 
and the Lorentz Invariance (LI)~\cite{Gehrmann:1999as} identities, 
which are implemented in the Mathematica based package LiteRed \cite{Lee:2013mka}.
Finally, we obtain 10 and 149 MIs at one- and two-loops, respectively.
The resulting set of MIs is systematically mapped on to those evaluated in \cite{Gehrmann:2013cxs,Gehrmann:2014bfa} as Laurent series 
in $\epsilon$ up to the required order.
Finally, substituting the results of MIs from \cite{Gehrmann:2013cxs,Gehrmann:2014bfa},
we obtain the two-loop result for the coefficient ${\mathcal{C}_2}$.  Both UV and IR divergences
appear as poles in $\epsilon$ at every order in $\hat a_s$.   In the next section, we demonstrate how
the renormalization of the strong and the Yukawa couplings render these coefficients UV finite leaving only
IR divergences.


\subsection{Ultraviolet renormalization} 
\label{renorm}
The scalar function $\mathcal{C}_2$ computed in powers of the bare coupling constant $\hat a_s$ contains both UV and IR
divergences.  Note that the entire amplitude is proportional to {the square of} $\hat {\lambda}_b$, the bare 
Yukawa coupling.   We use the modified minimal subtraction ($\overline{MS}$) scheme to perform the UV renormalization of the  
amplitudes.
In this scheme, the renormalized strong coupling constant $a_{s}$ 
is related to the bare strong coupling constant, $\hat{a}_{s}$, 
through the renormalization constant $Z\left(a_s(\mu_{R}^2),\epsilon\right)$ at the renormalization scale $\mu_{R}$ as
\begin{align} \label{UV1}
\frac{\hat{a}_{s}}{\mu_{0}^{\epsilon}} S_{\epsilon} = \frac{a_s}{\mu_{R}^{\epsilon}}Z\left(a_s(\mu_{R}^2),\epsilon\right),
\end{align}
where $Z\left(a_s(\mu_{R}^2),\epsilon\right)$ up to two-loops is given by
\begin{equation}
Z\left(a_s(\mu_{R}^2),\epsilon\right) = 1 + a_s\bigg(\frac{2\beta_{0}}{\epsilon}\bigg) 
             + a_s^2 \bigg( \frac{4\beta^{2}_{0}}{\epsilon^2} + \frac{\beta_{1}}{\epsilon}\bigg) 
             + \mathcal{O}(a_{s}^{3}) \, .
\end{equation} 
Here,
$S_{\epsilon} \equiv \exp[\left(\gamma_E-\ln4\pi\right)\frac{\epsilon}{2}]$
is the phase-space factor in $d$-dimensions, $\gamma_{E} = 0.5772...$ is the Euler-Mascheroni constant and $\mu_{0}$ is an arbitrary 
mass scale introduced to make $\hat g_s$ dimensionless in  $d$-dimensions. 
The constants $\beta_{0}$ and $\beta_{1}$ are the coefficients of $\beta$ function which, 
for $n_{f}$ light quark flavors, are found \cite{Gross:1973id,Politzer:1973fx,Caswell:1974gg,Jones:1974mm,Egorian:1978zx} to be
\begin{equation}
\beta_{0} = \frac{11}{3}C_{A} - \frac{4}{3}n_{f}T_{F},
\hspace{5mm} 
\beta_{1} = \frac{34}{3}C_{A}^{2} - \frac{20}{3}C_{A}n_{f}T_{F} - 4C_{F}n_{f}T_{F} \, ,
\end{equation}
where $C_{A} = N$, $C_{F} = (N^2 - 1)/2N$ are the Casimirs of SU(N) group and $T_{F} = 1/2$.
Similar to $\hat a_s$, the renormalization of the Yukawa coupling constant $\hat {\lambda}_{b}$ 
leads to renormalized $\lambda_{b}(\mu_{R}^2)$ at the renormalization scale $\mu_{R}$ through
\begin{align}\label{UV2}
\frac{\hat{\lambda}_{b}}{\mu_{0}^{\epsilon/2}}S_{\epsilon} 
&= \frac{\lambda_b}{\mu_{R}^{\epsilon/2}} Z_{\lambda}\left(a_s(\mu_{R}^2),\epsilon\right)
\nonumber\\
&= \frac{\lambda_b}{\mu_{R}^{\epsilon/2}} \left[1 + a_{s} \bigg(\frac{1}{\epsilon}Z^{(1)}_{\lambda,1}\bigg) 
 + a^2_{s}\bigg(\frac{1}{\epsilon^2}Z^{(2)}_{\lambda,2} + \frac{1}{\epsilon}Z^{(2)}_{\lambda,1}\bigg) + \mathcal{O}(a_{s}^3) \right],
\end{align}
where the coefficients $Z^{(i)}_{\lambda,j}$ are given by
\begin{equation}
 Z^{(1)}_{\lambda,1} = 6C_{F},\,Z^{(2)}_{\lambda,2} = 18C_{F}^2 + 6\beta_{0}C_{F},\, 
Z^{(2)}_{\lambda,1} = \frac{3}{2}C_{F}^2 + \frac{97}{6}C_{F}C_{A}-\frac{10}{3}C_{F}n_{f}T_{F}.
\end{equation}
The perturbative expansion of the amplitude for the aforementioned process 
in terms of the bare strong and Yukawa couplings is given by 
\begin{equation}\label{UV3}
\mathcal{A}_{ij} = \left(\frac{\hat{\lambda}_{b}}{\mu_{0}^{\epsilon/2}}S_{\epsilon}\right)^2\left[\hat{\mathcal{A}}^{(0)}_{ij}
+ \bigg(\frac{\hat{a}_s}{\mu_{0}^{\epsilon}}S_{\epsilon}\bigg) \hat{\mathcal{A}}^{(1)}_{ij} 
+ \bigg(\frac{\hat{a}_s}{\mu_{0}^{\epsilon}}S_{\epsilon}\bigg)^{2}\hat{\mathcal{A}}^{(2)}_{ij} 
+ \mathcal{O}(\hat{a}_{s}^{3})\right] \, ,
\end{equation}
where $\mathcal{\hat{A}}^{(l)}_{ij}$ is the $l^{th}$ loop unrenormalized amplitude. Similarly, the coefficient 
$\mathcal{C}_2$ replicates similar perturbative expansion of the following form,
\begin{equation}\label{UV4}
\mathcal{C}_{2} = \left(\frac{\hat{\lambda}_{b}}{\mu_{0}^{\epsilon/2}} S_{\epsilon} \right)^2 \left[\hat{\mathcal{C}}^{(0)}_{2} 
+ \bigg(\frac{\hat{a}_s}{\mu_{0}^{\epsilon}} S_{\epsilon} \bigg) \hat{\mathcal{C}}^{(1)}_{2}  
+ \bigg(\frac{\hat{a}_s}{\mu_{0}^{\epsilon}}S_{\epsilon}\bigg)^{2}\hat{\mathcal{C}}^{(2)}_{2}  
+ \mathcal{O}(\hat{a}_{s}^{3})\right] \, .
\end{equation}
In terms of the renormalized couplings, the coefficient $\mathcal{C}_2$ takes the form 
\begin{equation}\label{UV5}
\mathcal{C}_2 = \left(\frac{\lambda_b}{\mu_{R}^{\epsilon/2}}\right)^2\left[\mathcal{C}^{(0)}_2
+ a_s~\mathcal{C}^{(1)}_2 + a_s^2~\mathcal{C}^{(2)}_2 + \mathcal{O}(a_{s}^{3})\right] \, .
\end{equation}
We obtain the coefficients $\mathcal{C}_{2}^{(i)}$ using Eq.~(\ref{UV1}) and~(\ref{UV2}) in Eq.~(\ref{UV4}) and comparing with Eq.~(\ref{UV5}):
\begin{align}
&\mathcal{C}^{(0)}_2 = \hat{\mathcal{C}}^{(0)}_2 \, ,
\nonumber\\
&\mathcal{C}^{(1)}_2 = \frac{12}{\epsilon}C_F\hat{\mathcal{C}}^{(0)}_2 
                   + \frac{1}{\mu_{R}^{\epsilon}}\hat{\mathcal{C}}^{(1)}_2\, ,
\nonumber\\
&\mathcal{C}^{(2)}_2 = \left[\frac{12}{\epsilon^2}\Big(6C_{F}^2 
                   + \beta_{0}C_{F}\Big) +\frac{1}{\epsilon} 
                   \bigg( 3 C_{F}^2 + \frac{97}{3}C_{F}C_{A}-\frac{20}{3}C_{F}n_{f}T_{F} 
                   \bigg)\right]\hat{\mathcal{C}}^{(0)}_2 
\nonumber\\
&                   + \frac{2}{\mu_{R}^{\epsilon}}\left[\frac{\beta_0}{\epsilon} 
                   + \frac{6C_{F}}{\epsilon}\right]\hat{\mathcal{C}}^{(1)}_2 
                   + \frac{1}{\mu_{R}^{2\epsilon}}\hat{\mathcal{C}}^{(2)}_2 \, .
\end{align}
These constants $\mathcal{C}^{(l)}_{2}$, $l=0,1,2$,  
that result after performing the renormalization of the strong and the Yukawa couplings, 
are UV finite.  However they 
are sensitive to  both soft and collinear divergences which will be the topic of our next section.
These soft and collinear divergences show up in terms of poles in $\epsilon$.

\subsection{Infrared divergences and their factorization}
\label{infraredfact}
The UV finite amplitudes still contain divergences resulting from soft and collinear regions of  
the loop integrals.   They result from soft gluons and massless collinear quarks and gluons in the loops. 
In the physical observables, the soft and the collinear  divergences coming from the final states of the virtual diagrams cancel against 
those resulting from the phase space integrals of the real emission processes.
Due to the Kinoshita-Lee-Nauenberg (KLN) theorem \cite{Kinoshita:1962ur,Lee:1964is}, the cancellation takes
place order by order in perturbation theory.  While the soft divergences cancel fully,  
the collinear divergences resulting from initial massless states, do not cancel 
at the sub-process level.  Thanks to the collinear factorization theorem \cite{Collins:1985ue} 
these initial state  collinear divergences can be factored out in a process independent way and absorbed into the
bare parton distribution functions.  This procedure is called mass factorization which is
also a consequence of KLN theorem applied at the hadronic level.   
While all these IR divergences that appear in the amplitudes  
do not pose any problem for the physical observables,
they provide valuable information about the universal structure of the infrared divergences in the
QCD amplitudes.  In fact,  it can be shown that
these divergences systematically factor out from the amplitudes to all orders in perturbation theory \cite{Kidonakis:1998nf, Sen:1982bt}.
These factored IR divergences demonstrate the universal structure in terms of certain soft and collinear
anomalous dimensions.  An elegant proposal was put forth by Catani who predicted   
IR pole structure of the amplitudes up to two-loop level in non-abelian gauge theory 
\cite{Catani:1998bh}.  He demonstrated that the $n$-particle QCD amplitudes factorize in terms of
the universal IR subtraction operator denoted by $\mathcal{I}$.  This $\mathcal{I}$-operator has a dipole structure \cite{Catani:1998bh} containing
process independent universal cusp and collinear anomalous dimensions.  
Thanks to the wealth of results from two-loop calculations 
of the three-parton $q\bar{q}g$ amplitudes \cite{Garland:2001tf} 
and $2\rightarrow 2$ scattering amplitudes \cite{Anastasiou:2001sv,Glover:2001af,Bern:2002tk}, 
that involve non-trivial color structures \cite{Bern:2002tk,Bern:2004cz}, 
the $\mathcal{I}$-operator is completely known up to two-loop level. 
In \cite{Sterman:2002qn}, the authors provide further insight on the 
factorization and resummation properties of QCD amplitudes 
in the light of Catani's proposal and demonstrate a connection between divergences 
governed by soft and collinear anomalous dimensions, see also \cite{Aybat:2006wq,Aybat:2006mz}. 
There have been several efforts \cite{Becher:2009cu,Gardi:2009qi} to determine the structure of
$\mathcal{I}$-operator beyond two-loop level.
Following \cite{Catani:1998bh}  
we express one and two-loop UV renormalized amplitudes in terms of the
$\mathcal{I}$-operator as
\begin{align} \label{IR1}
\mathcal{C}^{(0)}_2(\epsilon) &= \mathcal{C}^{(0),\rm fin}_2(\epsilon) \, , \nonumber\\
\mathcal{C}^{(1)}_2(\epsilon)&= 2\mathcal{I}_{b}^{(1)} (\epsilon) \mathcal{C}^{(0)}_2 (\epsilon)
 + \mathcal{C}^{(1),\rm fin}_2 (\epsilon),
\nonumber\\
\mathcal{C}^{(2)}_2(\epsilon)&= 4\mathcal{I}_{b}^{(2)}(\epsilon)\mathcal{C}^{(0)}_2 (\epsilon)+2\mathcal{I}_{b}^{(1)} (\epsilon) \mathcal{C}^{(1)}_2 (\epsilon)
  + \mathcal{C}^{(2),\rm fin}_2 (\epsilon).
\end{align}
The matrix elements of the subtraction operator for the bottom quark, $\mathcal{I}_b$ are given by 
\begin{align}
\mathcal{I}_{b}^{(1)}(\epsilon) &= \frac{e^{-\frac{\epsilon}{2}\gamma_{E}}}{\Gamma\big(1+\epsilon/2\big)} \bigg(- \frac{4C_{F}}{\epsilon^2}
+\frac{3C_{F}}{\epsilon}\bigg) \bigg(-\frac{s}{\mu_{R}^2}\bigg)^{\frac{\epsilon}{2}},
\nonumber\\
\mathcal{I}_{b}^{(2)}(\epsilon) &= -\frac{1}{2}\mathcal{I}_{b}^{(1)}(\epsilon)\bigg(\mathcal{I}_{b}^{(1)}(\epsilon) 
- \frac{2\beta_{0}}{\epsilon}\bigg)
+ \frac{e^{\frac{\epsilon}{2}\gamma_{E}}\Gamma(1+\epsilon)}{\Gamma(1+\epsilon/2)}\bigg(-\frac{\beta_0}{\epsilon} 
+ K\bigg)\mathcal{I}_{b}^{(1)}(2\epsilon)
+ 2H_{b}^{(2)}(\epsilon),
\end{align}
with $K$ \cite{Catani:1998bh} and $H_{b}^{(2)}$ \cite{Sterman:2002qn} given by as follows
\begin{align}
K &= \left(\frac{67}{18}-\frac{\pi^2}{6}\right)C_{A} - \frac{10}{9}n_{f}T_{F},
\nonumber\\
H_{b}^{(2)} &= \left(-\frac{s}{\mu_{R}^2}\right)^{\epsilon}\frac{e^{-\frac{\epsilon}{2}\gamma_{E}}}{\Gamma\big(1+\epsilon/2\big)}
\frac{1}{\epsilon} \Bigg[C_{A}C_{F}\left(-\frac{245}{432} 
+ \frac{23}{16}\zeta_{2}-\frac{13}{4}\zeta_{3}\right) \nonumber\\
& \hspace{4cm} ~~  + C_{F}^{2}\left(\frac{3}{16}-\frac{3}{2}\zeta_{2} + 3\zeta_{3}\right)
+ C_{F}n_{f}\left(\frac{25}{216} - \frac{1}{8}\zeta_{2}\right)\Bigg] \, .
\end{align}
According to the proposal by Catani, the coefficients ${\cal C}_2^{(i),\rm fin} (\epsilon)$ should 
be free of IR divergences and hence are finite as $\epsilon \rightarrow 0$.
Since the resulting expression at two-loops level, ${\cal C}_2^{(2),\rm fin} (\epsilon)$ is quiet lengthy,
we had to simplify the expression first at the color factor level and then for each color factor,
terms of uniform transcendentality were further simplified.  We find that our final result
is in accordance with Catani's predictions for the IR poles, which serves as an important check on the 
correctness of our computation.
   
It is interesting to observe that the singlet contributions which are proportional to the color factor $C_F n_b T_F$,
for $n_b = 1$,
develops IR divergences at the intermediate stages of the computation. 
However at the end, all the IR singularities cancel among themselves
contributing only to the IR finite part.  This is consistent with the IR pole structure predicted by Catani.
The resulting finite constant ${\mathcal C}^{(2),\rm fin}_2$ that results after subtracting the
IR divergences using Catani's $\mathcal{I}$-operators is too lengthy to be presented here and hence  
attached as ancillary
file in Mathematica format.   


\section{Numerical Evaluation of the Amplitude}
\label{numerical}

The finite coefficients, $\mathcal{C}_2^{(i),\text{fin}}$, $i=1,2$,
obtained in Eq.~(\ref{IR1}) contain multiple classical polylogarithms, 
which are functions of the scaling variables $x$ and $y$. 
These polylogarithms can be attributed to different transcendental weights,
the property that we use to simplify the two loop coefficient, $\mathcal{C}_2^{(2),\text{fin}}$. 
Considering the complexity of our final result $\mathcal{C}_2^{(2),\text{fin}}$,
we perform a numerical evaluation using Mathematica for a wide range of 
scaling variables.   
\begin{figure}[h!]
\includegraphics[width=15cm,height=10cm]{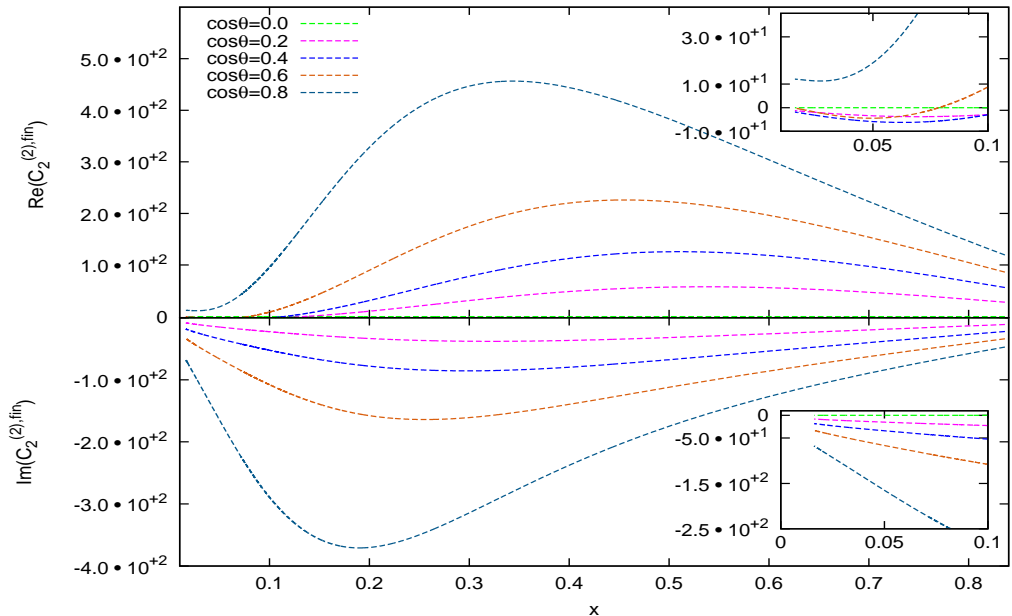}
\caption{\sf Behaviour of the real and the imaginary part of $\mathcal{C}_2^{(2),\rm fin}$ as a function of the scaling variable $x$
for different values of $\cos(\theta)$. The insets show the region close to $x=0$.}
\label{amp}
\end{figure}
More precisely, in Fig.~\ref{amp} we 
 plot the real and the imaginary parts of the coefficient $\mathcal{C}_2^{(2),\rm fin}$ 
as functions of the scaling variable $x$ for different values
of $\cos(\theta)$, where $\theta$ denotes the angle between one of the initial state 
fermions and the Higgs boson in the center of mass frame of incoming states. 
We consider $m_h = 125\, \text{GeV}$ and the renormalization scale as
$\mu_R^2=m_h^2/2$. In addition, we normalize the coefficient with the factor $m_h^2$. 
The amplitude is anti-symmetric under $\cos(\theta) \rightarrow - \cos(\theta)$, as expected for a purely fermionic amplitude. Since this symmetry has not been used in the setup of the calculation, it serves as a strong check on our results.
 Our expression contains  
polylogarithms that are multiplied by large rational coefficients, hence we encounter 
numerical instabilities during the evaluation.  To avoid this, we  evaluate 
the polylogarithms at double precision while setting the rational coefficients
at higher precision.  
From the Fig.~\ref{amp}, we observe a stable behaviour for the real and imaginary parts
for the range of parameters considered.
In addition, the dependence of the coefficient near the phase space
boundary $x=0$ is displayed in the insets. The simplified analytical 
results and their numerical implementation are provided
as separate ancillary files.

\section{Inclusive Cross Section up to NNLO}
\label{inclusive}
In this section, we describe in detail the computation of inclusive cross section
up to NNLO level for producing a pair of Higgs bosons resulting class-A and class-B diagrams.
The hadronic cross section can be expressed 
in terms of partonic cross sections appropriately convoluted with   
the corresponding bare parton distribution functions $\hat f_{a_i}(x_i), i=1,2$ as
\begin{eqnarray}
\label{hcross}
\sigma^{HH} = \sum_{a_1,a_2}\int dx_1 \hat f_{a_1} (x_1) \int dx_2 \hat f_{a_2}(x_2)  \hat 
\sigma^{HH}_{a_1 a_2}(x_1,x_2,m_h^2)\,,
\end{eqnarray}
where $x_i$ are the momentum fractions of initial state partons and $a_{1,2}=q ,\overline q,g$. 
$\hat \sigma^{HH}_{a_1 a_2}$ is the UV finite partonic cross section
for producing a pair of Higgs bosons along with 
$n_X$ number of colored particles (partons) through the reactions $a_1(p_1) + a_2(p_2) 
\rightarrow H(q_1) + H(q_2) + X({k_c})$ and is obtained using
\begin{eqnarray}
\label{pcross}
\hat \sigma^{HH}_{a_1 a_2} = {1 \over 2 s} \prod_{n=1}^2 \int d\phi(q_n)  \prod_{c=1}^{n_X}
\int d\phi(k_c)  \overline \sum |\mathcal{M}_{a_1 a_2}|^2 (2 \pi)^d \delta^d\Big(p_1+p_2 - \sum_{n=1}^2 q_n - \sum_{c=1}^{n_X} 
k_c\Big)
\end{eqnarray}
where $p_i,q_i$ and $k_c$ are the momenta of incoming partons, final state Higgs bosons and
partons respectively. 
In $d$-dimensions, the phase space measure $d \phi(p)$ of a final state particle with momentum $p$ and mass $m$ 
is given by
\begin{eqnarray}
\label{ps}
d\phi(p) = { d ^{d-1} \vec p \over (2 \pi)^{d-1} 2 p^0}
\end{eqnarray}
where $p^0 = \sqrt{m^2 +\left| \vec p\right|^2}$.  
$\mathcal{M}_{a_1 a_2}$ is the amplitude for the process $a_1(p_1) + a_2(p_2)
\rightarrow H(q_1) + H(q_2) + X({k_c})$ and is calculable
order by order in perturbative QCD.  
The symbol $\overline \sum$ indicates
that we have to sum over all the quantum numbers of final states, average over initial states 
and finally include the symmetry factor for final state identical particles.  
For convenience, we classify the partonic channels that contribute to $\mathcal{M}_{a_1 a_2}$ 
into class-A and class-B.   We find that these channels do not interfere
for the inclusive cross section as well as for the invariant mass distribution of Higgs boson pairs. 
Hence, in the following we treat them separately.

For the class-A diagrams, the amplitude $\mathcal{M}_{a_1 a_2}$ factorises 
into a product of two sub amplitudes, where one of them describes the production
of a single Higgs boson with virtuality $Q^2$ and the other encapsulates its decay to 
a pair of on-shell Higgs bosons.  By suitably factorising the phase space 
we can describe the entire reaction as a continuous process of
producing a single off-shell boson with different virtualities, subsequently 
decaying to a pair of on-shell Higgs bosons. 
In other words, we can write $\hat \sigma^{HH}_{a_1a_2}$ for class-A diagrams as
\begin{eqnarray}
\label{schan}
\hat \sigma^{HH}_{A,a_1 a_2} &=& \int {d Q^2 \over 2 \pi} 
\hat \sigma_{A,a_1a_2}^{H^*}(x_1,x_2,Q^2)
\left|{\cal P}_H(Q^2)\right|^2
2 Q \Gamma_A^{H^*\rightarrow HH}(Q^2)
\end{eqnarray}
where the ${\cal P}_H(Q^2)$ is the Higgs boson propagator, given by 
\begin{eqnarray}
{\cal P}_H(Q^2) = {i \over Q^2 - m_h^2 + i \Gamma_h m_h}
\end{eqnarray}
with $\Gamma_h$, the decay width of the Higgs boson.  The cross section that describes
the production of a Higgs boson with virtuality $Q^2$ is given by 
\begin{eqnarray}
\label{prod}
\hat \sigma_{A,a_1a_2}^{H^*}(x_1,x_2,Q^2) &=&
{1 \over 2  s} 
\prod_{c=1}^{n_X}  \int d\phi(k_c) \int d\phi(Q) \overline \sum |\mathcal{M}_{A,{a_1 a_2}}^{H^*}|^2
(2 \pi)^d \delta^d\Big(p_1+p_2 - Q - \sum_{c=1}^{n_X} k_c\Big)\,.
\nonumber
\\
\end{eqnarray}
Here $\mathcal{M}_{A,a_1 a_2}^{H^*}$ is the amplitude for the production of an off-shell 
Higgs boson with the virtuality $Q^2$ and $n_X$ number of colored particles.
The decay rate $\Gamma_A^{H^*\rightarrow HH}$ is given by 
\begin{eqnarray}
\label{dec}
\Gamma_A^{H^*\rightarrow HH}(Q^2) &= & \frac{1}{2Q}\prod_{n=1}^2 \int d\phi(q_n)   \overline \sum \left|\mathcal{M}_A^{H^*\rightarrow HH}\right|^2
(2 \pi)^d \delta^d\Big(Q - \sum_{n=1}^{2} q_n\Big) \,,
\end{eqnarray}
with $\mathcal{M}_{A}^{H^*\rightarrow HH}$ describing its decay into a pair of on-shell Higgs bosons.
The decay rate $\Gamma_{A}^{H^*\rightarrow HH}$ is straightforward to compute and in 4-dimensions
it is found to be 
\begin{eqnarray}
\label{decresult}
\Gamma_{A}^{H^*\rightarrow HH} (Q^2) = {9 \beta(Q^2) m_h^4 \over 32 \pi v^2 Q},
\quad \quad \quad  \beta(Q^2) = \sqrt{1 - {4 m_h^2 \over Q^2}}\,.
\end{eqnarray}
Substituting Eq.~(\ref{schan}) in Eq.~(\ref{hcross}) and using Eqs.~(\ref{prod},~\ref{dec}), we obtain
$\sigma_A^{HH}$, the contribution of class-A diagrams to $\sigma^{HH}$ in Eq.~(\ref{pcross}): 
\begin{eqnarray}
\label{sigHH}
\sigma_A^{HH} = \int {d Q^2 \over 2 \pi}  
D_H(Q^2) \sigma^{H^*}_A(Q^2)
\end{eqnarray}
with
\begin{eqnarray}
\label{singH}
\sigma^{H^*}_A(Q^2) &=& \sum_{a_1,a_2}\int dx_1 \hat f_{a_1} (x_1) \int dx_2 \hat f_{a_2}(x_2)  
\hat \sigma^{H^*}_{A,a_1 a_2}(z,Q^2)
\end{eqnarray}
where the partonic scaling variable $z=Q^2/s$ and $D_H(Q^2)= 2 Q \Gamma_A^{H^*\rightarrow HH}(Q^2) \left|{\cal P}_H(Q^2)\right|^2$.
Note that $\sigma^{H^*}_A$ is known exactly up to NNLO level \cite{Harlander:2003ai} and N$^3$LO level
\cite{Ravindran:2006cg,Ahmed:2014cha} in the soft plus virtual approximation
for on-shell production of single Higgs boson.  Hence, following \cite{Harlander:2003ai}, 
we can express $\sigma^{H^*}_A(Q^2)$ in terms of IR finite coefficients convoluted with renormalized parton distribution functions
$f_c(x,\mu_F)$ as   
\begin{eqnarray}
\label{sinHR}
\sigma^{H^*}_A(Q^2) &=& \sigma^{H^*}_0(Q^2,\mu_R)
\sum_{a_1,a_2}\int dx_1 f_{a_1} (x_1,\mu_F) \int dx_2 f_{a_2}(x_2,\mu_F)  
z \Delta_{A,a_1 a_2}(z,Q^2,\mu_F,\mu_R)
\nonumber \\
\end{eqnarray}
where $\sigma^{H^*}_0(Q^2,\mu_R)= {\pi m_b^2(\mu_R^2) /(6 Q^2 v^2)}$.  
$\Delta_{A,a_1 a_2}$ can be expanded in powers of strong coupling constant as
\begin{eqnarray}
\Delta_{A,a_1 a_2} (z,Q^2,\mu_F,\mu_R) = \sum_{i=0}^\infty a_s^i(\mu_R^2) \Delta_{A,a_1 a_2}^{(i)}(z,Q^2,\mu_F,\mu_R)\,.
\end{eqnarray}
Substituting Eq.~(\ref{sinHR}) in Eq.~(\ref{sigHH}) and making suitable change of variables, we obtain 
\begin{eqnarray}
\label{phenoA}
\sigma_A^{HH} = \sum_{a_1 a_2}\int_{\tau}^1 {d x \over 2 \pi } 
\Phi_{a_1 a_2} (x,\mu_F) 
\int_{\tau \over x}^1 dz 
\left[\sigma_0^{H^*}(Q^2,\mu_R) D_H(Q^2) 
\Delta_{A,a_1 a_2} (z,Q^2,\mu_F,\mu_R) \right]_{Q^2 = x z S}
\end{eqnarray}
where $\tau = 4 m_h^2/S$, $S=s/x_1 x_2 $, the hadronic center of mass energy of
incoming hadrons and 
the partonic flux $\Phi_{a_1 a_2}(x,\mu_F)$ is given by
\begin{eqnarray}
\Phi_{a_1 a_2}(x,\mu_F) = \int_x^1 {dy \over y} f_{a_1}(y,\mu_F) f_{a_2}\left({x \over y},\mu_F\right)\,.
\end{eqnarray}
In the next section, we use Eq.~(\ref{phenoA}) to obtain the numerical impact of class-A diagrams
to the inclusive production cross section.

We now describe how the contributions to Eq.~(\ref{hcross}) from class-B diagrams can be obtained.
Since class-B diagrams contain, in addition, t and u channels,  the corresponding amplitudes 
do not factorise like
class-A diagrams, making the computation technically more challenging beyond NLO level.  
However, one can obtain certain dominant contributions of 
class-B processes resulting from soft gluon emission as they are process independent. 
Using the contributions from soft gluons, obtained in \cite{Ravindran:2006cg}, and those 
from the two-loop virtual processes 
computed in the previous sections, we can readily calculate the soft plus virtual contribution 
up to NNLO level, a first step towards obtaining the full NNLO contribution from class-B.  

For the class-B, the leading order contribution results from the Born 
process $b + \overline b \rightarrow  H +H $ 
contain $t$ and $u$ channels.
At NLO, one loop virtual corrections to Born and real emission processes
$b + \overline b \rightarrow H+H+g$ and $b (\overline b) +g \rightarrow H+H+b(\overline b)$
contribute.  The UV divergences that are present in the virtual processes 
to Born processes are removed using $\overline {MS}$ renormalisation scheme.  The soft 
and final state collinear divergences  
in both virtual as well as real emission processes cancel among each other
while the initial state collinear divergences are factored out and 
absorbed into bare bottom quark densities 
in the $\overline{MS}$ scheme through the mass factorization.
For the sub-process $b (\overline b) + g \rightarrow H+H+b(\overline b)$,
we encounter only collinear divergences and they are removed by mass factorization.   
We achieve this by using the semi analytical method, namely the 
two cut off phase space slicing \cite{Harris:2001sx}. 
The first computation of
NLO correction to production of a pair of Higgs bosons in bottom quark annihilation process
was achieved in \cite{Dawson:2006dm} by using the same approach. In the present article, we  
use this approach only for the class-B diagrams to compute NLO contributions.  In this method, 
for $b +\overline b \rightarrow H+H+g$, 
two slicing parameters $\delta_s$ and $\delta_c$
are introduced to separate three body phase space into soft, hard collinear and hard non-collinear
regions, while for $g+b (\overline b) \rightarrow H+H+ b(\overline b)$, we need to introduce
only $\delta_c$ as these are free from soft divergences. 
The slicing parameter $\delta_s$ divides the real emission phase space into soft and hard regions.
The soft region is the part of the phase space where the energy of the gluon in the center-of-mass frame of
incoming partons is required to be less than $\delta_s \sqrt{s}/2$ and the rest is called hard region.
The later contains collinear configurations where the two massless partons become collinear to
each other leading to collinear singularities.  We use $\delta_c$ to divide the hard
region into hard-collinear and 
hard non-collinear regions, denoted respectively by $\mathcal{HC}$ and $\overline{\mathcal{HC}}$.  
Keeping these slicing parameters $\delta_s$ and $\delta_c$ infinitesimally small, 
the virtual loop integrals and 
the soft and collinear sensitive phase space integrals are computed within  
the method of dimensional regularization.
The corresponding singularities show up as poles in dimensional regularization parameter $\epsilon$.

We describe below the essential steps that are followed in dealing with IR singularities 
in phase space slicing method.  We start with UV finite hadronic cross section at NLO level,
denoted by $d \sigma^{HH+1}$.   It gets contribution from real emission 
partonic sub-process  $a_1 + a_2 \rightarrow HH + a_3$ where 
the final state consists of a pair of Higgs bosons $HH$ and $a_3$, a single partonic state. 
We divide the phase space of $a_3$ into three regions using two slicing parameters as
\begin{eqnarray}
d \sigma^{HH+1}(\delta_s,\delta_c,\epsilon) = 
d \sigma^{HH,\mathcal{S}}(\delta_s,\epsilon) + 
d \sigma^{HH,\mathcal{HC}}(\delta_s,\delta_c,\epsilon)+ 
d  \sigma^{HH, \overline {\mathcal{HC}}}(\delta_s,\delta_c) \,.
\end{eqnarray}
The soft ($d \sigma^{HH,\mathcal{S}}(\delta_s,\epsilon)) $ and hard-collinear  
($d  \sigma^{HH,\mathcal{HC}}(\delta_s,\delta_c,\epsilon) )$ contributions can be computed analytically when the slicing
parameters are infinitesimally small within the dimensional regularization.  
Soft and collinear singularities appear as poles in $\epsilon$ and are cancelled against
those resulting from the virtual diagrams as well as from the counter terms that are used to 
perform mass factorization.
In other words, the following sum, denoted by $ d \sigma^{HH}_{\text{NLO}}$  is finite as 
$\epsilon \rightarrow 0$:
\begin{eqnarray}
d \sigma^{HH}_{\text{NLO}}(\mu_F) = 
d \sigma^{HH,\mathcal{V}}(\epsilon) + 
d \sigma^{HH+1}(\delta_s,\delta_c,\epsilon) + 
d \sigma^{HH,\text{CT}}(\delta_s,\delta_c,\epsilon,\mu_F) 
\end{eqnarray}
where $d  \sigma^{HH,\mathcal{V}} (\epsilon)$ is the contribution from virtual corrections to Born level processes.
The counter term $d \sigma^{HH,\text{CT}}(\delta_s,\delta_c,\epsilon,\mu_F)$ that 
removes the initial state collinear singularities is defined at the factorization scale
$\mu_F$.  
While the sum given by 
\begin{eqnarray}
d \sigma^{\mathcal{S}+\mathcal{V}+\mathcal{HC}+\text{CT}}(\delta_s,\delta_c,\mu_F) &=&
d \sigma^{HH,\mathcal{S}}(\delta_s,\epsilon) + 
d \sigma^{HH,\mathcal{V}}(\epsilon) 
\nonumber\\
&& + d \sigma^{HH,\mathcal{HC}}(\delta_s,\delta_c,\epsilon)+ 
d \sigma^{HH,\text{CT}}(\delta_s,\delta_c,\epsilon,\mu_F) \,.
\end{eqnarray}
is free from soft and collinear poles in $\epsilon$, it 
depends on the slicing parameters.   
However, when the above sum is added to the hard non-collinear contributions ($d  
\sigma^{HH,\overline{\mathcal{HC}}}$), that is,
\begin{eqnarray}
\label{PSsum}
d \sigma^{HH}_{\text{NLO}} (\mu_F) = \lim_{\delta_s,\delta_c \rightarrow 0}\left( d  \sigma^{\mathcal{S}+\mathcal{V}+\mathcal{HC}+\text{CT}} (\delta_s,\delta_c) + d  \sigma^{HH,\overline{\mathcal{HC}}}(\delta_s,\delta_c)   \right)
\end{eqnarray}
the resulting contribution, Eq.~(\ref{PSsum}),
is guaranteed to be independent of the slicing parameters in the limit when they 
are taken to be infinitesimally small.
For the sub-process, $g + b (\overline b) \rightarrow H + H + b (\overline b)$, we encounter only collinear
divergences and hence we require a single slicing parameter $\delta_c$ to obtain infrared safe observable.

For completeness, we present the individual contributions that are required in phase space
slicing method to obtain inclusive cross section up to NLO level from class B diagrams.
 The virtual contribution for the sub-process initiated by $b$ and $\bar{b}$ is found to be
\bea
\label{xvirt}
d  \s^{HH,\mathcal{V}} &=& a_s(\muf^2) \left( \frac{s}{\muf^2} \right)^{\frac{\displaystyle{\e}}{2}}  \frac{\Gamma(1+\frac{\epsilon}{2})}{\Gamma(1+\epsilon)}dx_1dx_2
              \Bigg[C_F \left(-\frac{16}{\e^2} + \frac{12}{\e} \right)
\nonumber \\[2ex]
           && \times
                d \s^{HH,(0)}_{b \overline b}(x_1,x_2,\e)  \Big( f_{b}(x_1)f_{\overline b}(x_2) + \left(x_1 \leftrightarrow x_2 \right)\Big)
\nonumber \\[2ex]
&& +  d \s^{HH,\mathcal{V}}_{b \overline b, \text{fin}}(x_1,x_2,\e)  \Big( f_{b}(x_1)f_{\overline b}(x_2) 
+ \left(x_1 \leftrightarrow x_2\right) \Big)
\Bigg]
\eea
after setting renormalization scale $\mu_R=\mu_F$. 
The finite part of the virtual corrections, $d \sigma^{HH,\mathcal{V}}_{b \overline b,\text{fin}}$ can be obtained 
in terms of ${\cal C}_2$ given in Eq.~(\ref{C1C2}).
The soft contribution is given by
\bea
\label{xsoft}
d  \s ^{HH,\mathcal{S}} &\simeq&  a_s(\muf^2)  \left( \frac{s}{\muf^2} \right)^{\frac{\displaystyle{\e}}{2}}  \frac{\Gamma(1+\frac{\epsilon}{2})}{\Gamma(1+\epsilon)}
C_F ~\left(\frac{16}{\e^2} + \frac{16 \ln \ds}{\e} + 8 \ln^2 \ds \right)
\nonumber \\ [2ex]
&& \times \left( ~ d \s^{HH,(0)}_{b \overline b}(x_1,x_2,\e) f_{b}(x_1)f_{\overline b}(x_2
)
+ \left(x_1 \leftrightarrow x_2\right) \right)dx_1 dx_2\,.
\eea
The sum of hard-collinear and counter term contributions from both $b\bar{b}$ annihilation
and $gb(\bar{b})$ scattering processes, is found to be:
\bea
\label{xcoll}
\!\!\!\!\!\!  d \sigma^{\mathcal{HC} + \text{CT}}  &=&  a_s(\muf^2)  \left( \frac{s}{\muf^2} \right)^{\frac{\displaystyle{\e}}{2}}  \frac{\Gamma(1+\frac{\epsilon}{2})}{\Gamma(1+\epsilon)} dx_1 dx_2
\nonumber\\[2ex]
&&  \times \Bigg[d{ \sigma}^{HH,(0)}_{b\overline b}(x_1,x_2,\e) \Bigg\{ \frac{1}{2} f_{\overline b}(x_1,\mu_F) {\tilde f}_{b}(x_2,\mu_F) +
               \frac{1}{2} {\tilde f}_{\overline b}(x_1,\muf)f_{b}(x_2,\mu_F)
\nonumber\\[2ex]
&&           + 2 \left( -\frac{1}{\e} +\frac{1}{2} \ln \frac{ s}{\muf^2} \right) A_{b \rarrow b + g}
              ~  f_{\overline b}(x_1,\muf) f_{b}(x_2,\muf) + \left(x_1 \leftrightarrow x_2\right)  \Bigg\}
\Bigg].
\eea

Using the diagonal Altarelli-Parisi (AP) splitting function $P_{bb}(z)$, 
we find
\bea
A _{b \rarrow b+g} \equiv \int_{1- \delta_s}^1 \frac{dz}{z} P_{bb}(z) &=& 4C_F \left(2 \ln \ds + \frac{3}{2} \right) ,
\eea
and from the non-diagonal ones, we obtain
\bea
{\tilde f}_{b}(x, \mu_F)
    &=&  \int_x^{1 -\delta_s} \frac{dz}{z} f_b \left( \frac{x}{z}, \mu_F \right) {\tilde P}_{bb}(z)
      + \int_x^{1}            \frac{dz}{z} f_g \left( \frac{x}{z}, \mu_F \right) {\tilde P}_{bg}(z) ,
\eea
with
\be
{\tilde P}_{ij}(z) = P_{ij}(z) \ln \left( \delta_c \frac{1-z}{z} \frac{ s}{\mu_F^2} \right) +2 P^{\prime}_{ij}(z)\,,
\ee
where $P_{ij}^\prime (z)$ \cite{Harris:2001sx} are $\epsilon$ dependent part of AP splitting functions, that is 
\be
P_{ij}(z,\e) = P_{ij}(z)+\e P_{ij}^\prime (z)\,.
\ee

Adding all the order $a_s$ pieces together: the virtual cross-section $d \sigma^{HH,\mathcal{V}}$
in Eq.~(\ref{xvirt}), the
soft piece $d \sigma^{HH,\mathcal{S}}$ in Eq.~(\ref{xsoft}) and the mass factorized hard-collinear contribution
$d\sigma^{HH,\mathcal{HC}+\text{CT}}$
as given in Eq.~(\ref{xcoll}),
we find that the poles in $\e$ cancel in the sum given in Eq.~(\ref{PSsum}) giving IR finite 
NLO contribution from class-B diagrams.  In the next section, we use 
this semi-analytical result to study the numerical impact of class-B NLO contributions 
on the inclusive cross section.

Going beyond NLO for the class-B diagrams requires a dedicated computation taking into account
pure virtual contributions presented in the present work, the double real and 
single real-virtual contributions.  The inclusion of the later contributions
is beyond the scope of this present work.  However,
we can compute the SV contribution resulting from class-B diagrams
using the two-loop virtual contributions computed in the present article and 
the universal soft and collinear
parts.  To achieve this, we follow the general formalism presented in~\cite{Ravindran:2006cg}, which is applicable to both classes of diagrams.  

We begin with the UV finite partonic cross section for producing a pair of Higgs bosons and 
$n_X$ partons, namely for the process $b(p_1) + \overline b(p_2) \rightarrow H(q_1) + H(q_2) + X(k_c)$,
\begin{eqnarray}
\hat \sigma_{b\overline b} = {1 \over 2  s} \prod_{n=1}^2 \int d\phi(q_n)  \prod_{c=1}^{n_X} 
\int d\phi(k_c)\overline \sum |\mathcal{M}_{b \overline b}|^2  (2 \pi)^d \delta^d\Big(p_1+p_2 - \sum_{n=1}^2 q_n - \sum_{c=1}^{n_X} 
k_c\Big) 
\end{eqnarray}
where $c$ counts the number of partons in the final state. 
The dominant soft gluon contributions to partonic reactions  
are proportional to terms such as  $\delta(1-z)$ and $+$ distributions:
\begin{eqnarray}
{\cal D}_j(z) =\left( {\log^j(1-z) \over 1-z } \right)_+ .
\end{eqnarray}
Such contributions result only from bottom quark annihilation sub processes.
They themselves do not constitute infrared safe observables until
we include pure virtual contributions and mass factorization counter-terms. 
The resulting one is called SV contribution.

We briefly discuss how these contributions can be obtained.
In the soft limit, it is well known that 
the square of the real emission partonic matrix elements factorises into hard and soft parts 
and similarly
the phase space splits into their respective phase spaces.  The soft part when combined with
the pure virtual corrections and the mass factorization counter terms, will give infrared safe 
SV part of the cross section:  
\begin{eqnarray}
\hat \sigma^{\text{SV}}_{b\overline b} &=& \int {d Q^2 \over Q^2} {1 \over 2  s} \prod_{n=1}^2 
\int d\phi(q_n) \overline \sum |\mathcal{M}^{(0)}_{b \overline b}|^2
(2 \pi)^d \delta^d\Big(p_1+p_2 - \sum_n q_n\Big)  
\nonumber\\
&& \times 
\int d \phi(Q) \prod_{c=1}^{n_X} \int d \phi(k_c)\overline \sum |\mathcal{M}^{\text{SV}}|^2
(2 \pi)^d \delta^d\left(p_1+p_2 - Q - \sum_c k_c\right) 
\end{eqnarray}
where $\mathcal{M}^{(0)}_{b \overline b}$ is the Born amplitude for producing a pair of Higgs bosons in bottom quark annihilation
and $\mathcal{M}^{\text{SV}}$ is the SV part of amplitude $\mathcal{M}_{b \overline b}$.   
The second line of the above equation can be computed order by order in perturbation theory
for any colorless state with momentum $Q$ in a process independent way as the amplitude for the production
of a pair of Higgs bosons factorises out at every order.
Beyond the leading order,  virtual corrections to Born amplitudes and multiple soft gluon emissions  
both from tree level amplitudes as well as from the loop corrected amplitudes contribute to 
the SV.  While the singularities from soft gluons cancel between real and virtual amplitudes, 
the initial state collinear singularities can be removed only after adding appropriate 
mass factorization counter terms computed in the soft limit at the factorization scale $\mu_F$.      
The resulting hadronic cross section will be free of soft and collinear singularities:
\begin{eqnarray}
\sigma^{HH,\text{SV}} &=& \int {d Q^2 \over Q^2} 
\sum_{b,\overline b} \int dx_1 f_b(x_1,\mu_F) \int dx_2 f_{\overline b} (x_2,\mu_F)  
{1 \over 2  s} \prod_{n=1}^2 \int d \phi(q_n) 
\nonumber\\
&&
\times (2 \pi)^d \delta^d\Big(p_1+p_2-\sum_{n=1}^2 q_n\Big)
\overline \sum \sum_{i=A,B} \Delta_{i,b\overline b}^{\text{SV}} 
\left(\{p_j\cdot q_k\},z,Q^2,\mu_F,\mu_R\right)
\end{eqnarray}
where $z = Q^2/ s$, $i$ runs over both the classes of diagrams. 
Following \cite{Ravindran:2006cg}, the finite coefficients $\Delta_{i,b\overline b}^{\text{SV}}$ can be computed 
order by order in perturbation theory using one and two-loop virtual amplitudes, soft distribution function and diagonal mass factorization kernels. 
We expand $ \Delta_{i,b \overline b}^{\text{SV}}$ in powers of strong coupling constant as,
\begin{eqnarray}
\Delta_{i,b \overline b}^{\text{SV}} = \sum_{j=0}^\infty a_s^j(Q^2) \Delta_{i,b \overline b}^{\text{SV}, (j)}(Q^2)
\end{eqnarray}
where we have set $\mur^2=\muf^2 = Q^2$. 
The coefficients, $\Delta_{i,b \overline b}^{\text{SV},(j)}$ for $j = 0,1,2$ can be expressed in terms of  
the cusp $A^q_{j}$, the soft 
$f^q_{j}$ and the collinear $B^q_{j}$ anomalous dimensions that are present in the virtual amplitudes and in 
the soft distribution function \cite{Ravindran:2006cg}:
\begin{eqnarray}
   \Delta^{\text{SV},(0)}_{i,b \overline b} &=& \delta(1-z)  |{\cal A}^{(0)}_{i,0}|^2\,,
\nonumber\\
   \Delta^{\text{SV},(1)}_{i,b \overline b} &=& \delta(1-z) \left\{  |{\cal A}^{(0)}_{i,0}|^2    \left( 2 
          {\cal \overline G}_1^{q,(1)}  \right)
       + {\cal A}^{(1)}_{i,0}   {\cal A}^{\star(0)}_{i,0}  
       +  {\cal A}^{(0)}_{i,0}   {\cal A}^{\star(1)}_ {i,0}\right\}
       + {\cal D}_0  |{\cal A}^{(0)}_{i,0}|^2    \left(  - 2  f_1^{q}  \right)
\nonumber\\&&
       + {\cal D}_1  |{\cal A}^{(0)}_{i,0}|^2    \left( 4  A_1^{q}  \right)\,,
\nonumber\\
\Delta^{\text{SV},(2)}_{i,b \overline b}   &=& \delta(1-z) \Bigg\{  |{\cal A}^{(1)}_{i,0}|^2    
       +   |{\cal A}^{(0)}_{i,0}|^2    \Big( 
          {\cal \overline G}_1^{q,(2)}  + 2 ({\cal \overline G}_1^{q,(1)})^2 + 2 
         \beta_0  {\cal \overline G}_2^{q,(1)}  - 8 \zeta_3  A_1^{q} f_1^{q}  
\nonumber\\ &&
- 2 \zeta_2  (f_1^{q})^2 - {4 \over 5} \zeta_2^2 
          (A_1^{q})^2 \Big)
       +    {\cal A}^{(2)}_{i,0}   {\cal A}^{\star(0)}_{i,0}    
       +   {\cal A}^{(1)}_{i,2}   {\cal A}^{\star(1)}_{i,-2}    
\nonumber\\ &&
       +    {\cal A}^{(1)}_{i,2}   {\cal A}^{\star(0)}_{i,0}    \Big( 4  A_1^{q}  \Big)
       +  {\cal A}^{(1)}_{i,1}   {\cal A}^{\star(1)}_{i,-1}    
       +   {\cal A}^{(1)}_{i,1}   {\cal A}^{\star(0)}_{i,0}    \Big(  - 2  f_1^{q}  - 4  B_1^{q}  \Big)
       +   {\cal A}^{(1)}_{i,-1}   {\cal A}^{\star(1)}_{i,1}    
\nonumber\\&&
       +   {\cal A}^{(1)}_{i,-2}   {\cal A}^{\star(1)}_{i,2}   
       +     {\cal A}^{(1)}_{i,0}   {\cal A}^{\star(0)}_{i,0}    \Big( 2  {\cal \overline G}_1^{q,(1)}  \Big)
       +    {\cal A}^{(0)}_{i,0}   {\cal A}^{\star(2)}_{i,0}    
       +    {\cal A}^{(0)}_{i,0}   {\cal A}^{\star(1)}_{i,2}    \Big( 4  A_1^{q}\Big)  
\nonumber\\&&
       +    {\cal A}^{(0)}_{i,0}   {\cal A}^{\star(1)}_{i,1}    \Big(  - 2  f_1^{q}  - 4  B_1^{q}  \Big)
       +  {\cal A}^{(0)}_{i,0}   {\cal A}^{\star(1)}_{i,0}    \Big( 2  {\cal \overline G}_1^{q,(1)}  \Big) \Bigg\}
\nonumber\\&&       
       + {\cal D}_0 \Bigg\{ |{\cal A}^{(0)}_{i,0}|^2   \Big(  - 2  f_2^{q}  - 4 
          f_1^{q}   {\cal \overline G}_1^{q,(1)}  
      - 4 \beta_0 
          {\cal \overline G}_1^{q,(1)}  + 16 \zeta_3  (A_1^{q})^2 + 8 
         \zeta_2  A_1^{q}   f_1^{q}  \Big)
\nonumber\\&&         
       + {\cal A}^{(1)}_{i,0}   {\cal A}^{\star(0)}_{i,0} 
         \Big(  - 2  f_1^{q}  \Big)
       + {\cal A}^{(0)}_{i,0}   {\cal A}^{\star(1)}_{i,0} 
         \Big(  - 2  f_1^{q}  \Big)\Bigg\}
\nonumber\\&&
       + {\cal D}_1 \Bigg\{ |{\cal A}^{(0)}_{i,0}|^2    \Big( 4  (f_1^{q})^2 + 4 
          A_2^{q}  + 8  A_1^{q}   {\cal \overline G}_1^{q,(1)}  + 4 \beta_0
           f_1^{q}  - 16 \zeta_2  (A_1^{q})^2 \Big)
\nonumber\\&&
       +  {\cal A}^{(1)}_{i,0}   {\cal A}^{\star(0)}_{i,0} 
         \Big( 4  A_1^{q}  \Big)
       +  {\cal A}^{(0)}_{i,0}   {\cal A}^{\star(1)}_{i,0} 
         \Big( 4  A_1^{q}  \Big)\Bigg\}
\nonumber\\&&
       + {\cal D}_2  |{\cal A}^{(0)}_{i,0}|^2    \Bigg\{  - 12  A_1^{q}  
          f_1^{q}  - 4 \beta_0  A_1^{q}  \Bigg\}
       + {\cal D}_3  |{\cal A}^0_{i,0}|^2    \Big\{ 8  (A_1^{q})^2 \Big\}\,,
\end{eqnarray}
where $\zeta_2=1.64493407\cdots,\zeta_3=1.20205690\cdots$ and ${\cal A}_{i,k}^{(j)}$ are obtained from Eq.~(\ref{C1C2}) by defining
${\cal A}_{mn} = {\cal A} \delta_{mn}$ and expanding in powers of $\epsilon$ as
\begin{eqnarray}
{\cal A}_i^{(j)}(\epsilon) = \sum_{k=-2 j}^\infty \epsilon^k {\cal A}_{i,k}^{(j)}\,.
\end{eqnarray}
The cusp ($A^q_i$,$i=1,2$) and collinear anomalous dimensions ($B_q^1$) are given by
\begin{align}
 A^q_1 &= 4 C_F \,,
\nonumber \\
 A^q_2 &= 8 C_F C_A \Big\{ \frac{67}{18} - \zeta_2 \Big\} + 8 C_F n_f \Big\{ -\frac{5}{9} \Big\} \,,
\nonumber \\
 B^q_1 &= 3 C_F \,,
\end{align}
and the soft anomalous dimensions are
\begin{align}
 f_1^q &= 0 \,,
\nonumber \\
 f_2^q &= C_A C_F \Big\{ -\frac{22}{3} {\zeta_2} - 28 {\zeta_3} + \frac{808}{27} \Big\}
        + C_F n_f T_F \Big\{ \frac{8}{3} {\zeta_2} - \frac{224}{27} \Big\} \,.
\end{align} 
The universal constants ${\overline {\cal G}}^{q,(j)}_k$ are given by
\begin{align}
  {\overline {\cal G}}^{q,(1)}_1 &= C_F \left( - 3 \zeta_2 \right) \,,\nonumber \\
  {\overline {\cal G}}^{q,(1)}_2 &= C_F \left( \frac{7}{3} \zeta_3 \right) \,, \nonumber\\
%
%
  {\overline {\cal G}}^{q,(2)}_1 &=  C_F n_f  \left( - \frac{328}{81} + \frac{70}{9} \zeta_2 + \frac{32}{3} \zeta_3 \right)
             + C_A C_F  \left( \frac{2428}{81} - \frac{469}{9} \zeta_2
                       + 4 {\zeta_2}^2 - \frac{176}{3} \zeta_3 \right) \,. 
%
%
\end{align}
Finally, defining $\overline \Delta^{\text{SV}}_{b \overline b}(z,Q^2,\mu_F,\mu_R)$ by 
\begin{eqnarray}
\overline \Delta^{\text{SV}}_{b \overline b}(z,Q^2,\mu_F,\mu_R) &=&
{1 \over 2  s} \prod_{n=1}^2 \int d \phi(q_n) (2 \pi)^d \delta^d\left(p_1+p_2-\sum_{n=1}^2 q_n\right)
\nonumber\\
&& \times  \overline
\sum |\mathcal{M}^{(0)}_{b \overline b}|^2 \sum_{i=1}^2 \Delta_{i,b\overline b}^{\text{SV}} \left(\{p_j\cdot q_k\},z,Q^2,\mu_F,\mu_R\right)\,,
\end{eqnarray}
we obtain $\sigma^{HH,\text{SV}}$:  
\begin{eqnarray}
\sigma^{HH,\text{SV}} = \int_{\tau}^1 {d x }~ 
\Phi_{a_1 a_2} (x,\mu_F) 
\int_{\tau \over x}^1 dz~ 
\overline \Delta_{a_1 a_2}^{\text{SV}} (z,Q^2,\mu_F,\mu_R) \Big) \Bigg|_{Q^2 = x z S}\,.
\end{eqnarray}
We have used the above formula to study the numerical impact of SV part of 
the partonic cross section resulting from class-B diagrams up to NNLO level on the inclusive production
of a pair of Higgs bosons.  

\section{Phenomenology}
\label{pheno}
In this section, we present in detail the numerical impact of our analytical results obtained
in the previous sections.
We mainly focus on the inclusive cross section for producing a pair of Higgs bosons at the LHC with the
center-of-mass energy $\sqrt{S} = 14$ TeV.  We use MMHT2014(68cl) PDF set \cite{Harland-Lang:2014zoa} and the corresponding 
$a_s$ through the LHAPDF-6 \cite{Buckley:2014ana} interface at every order in perturbation theory.  We use 
the running bottom quark mass renormalized in ${\overline {MS}}$ (see Eq.(13) in \cite{Harlander:2003ai})  
scheme with the
boundary condition $\overline m_b(m_b) = 4.7$ GeV.  Both $a_s(\mu_R)$ and $m_b(\mu_R)$ at various orders in 
perturbation theory are evolved using appropriate QCD $\beta$-function coefficients and quark mass anomalous
dimensions.  Similarly, the PDFs are evolved to factorization scale $\mu_F$ using the splitting functions 
computed to desired accuracy in the perturbation theory.  We choose the Higgs boson mass $m_h=125$ GeV
and its total decay width $\Gamma_h = 0.001$ GeV.
In our analysis, we have included all the partonic channels upto NNLO
level for the class-A diagrams while for the class-B, 
we could do this only up to NLO level, however, at NNLO level we have included SV contributions.
We find that this approximation does not change our conclusion as the dominant contribution 
results from class-A. To illustrate this point we state some of our observations 
from our numerical results. We find that the LO contributions from class-A diagrams 
are three orders of magnitude larger than those from class-B diagrams. 
We also find that NLO contributions change the LO cross section by $-1.096\%$ and at the NNLO level the 
change is about $-8.095\%$. 
The numerical result manifests the fact that the SV contribution presented in this paper 
not only gets the dominant contribution from class-A 
but also the stability of our NNLO result for di-Higgs production 
from the $b\bar{b}$ annihilation channel. 
We find that the contribution from bottom quark annihilation processes 
is two orders of magnitude smaller than from the gluon fusion processes.  
However, former ones need to be included for the precision studies at the LHC.  

Having studied the size of the corrections both at NLO and NNLO level,  
it is important to quantify the uncertainties resulting  from
the mass scales introduced in our calculations. 
Recall that the renormalisation of the ultra-violet and the initial state collinear divergences 
enforces the introduction of mass scales namely $\mu_R$ and $\mu_F$ respectively.  
The $\mu_R$ dependency shows up in the coupling constant $a_s(\mu_R)$, the mass $m_b(\mu_R)$
and in the mass factorised partonic cross sections at 
various orders in perturbations theory.  The coupling constants are evolved using 
the appropriate QCD $\beta$-function coefficients and quark mass anomalous dimensions. 
The $\mu_F$ scale dependency comes from the PDFs 
that are evolved using splitting functions 
computed in the perturbation series. 
But the cross section, like every other physical observables, is expected to be 
independent of these arbitrary mass scales.  This crude fact manifests 
the scale independency if we sum the perturbative predictions to all orders in perturbation theory. 
Since we have truncated the series, there is a residual scale dependency.  
In the following we aim to study this 
by varying both $\mu_R$ and $\mu_F$ scales.
\begin{figure}
\hspace{-2cm}
\centering
\begin{subfigure}{.5\textwidth}
 \centering
  \includegraphics[width=.75\linewidth, angle=270]{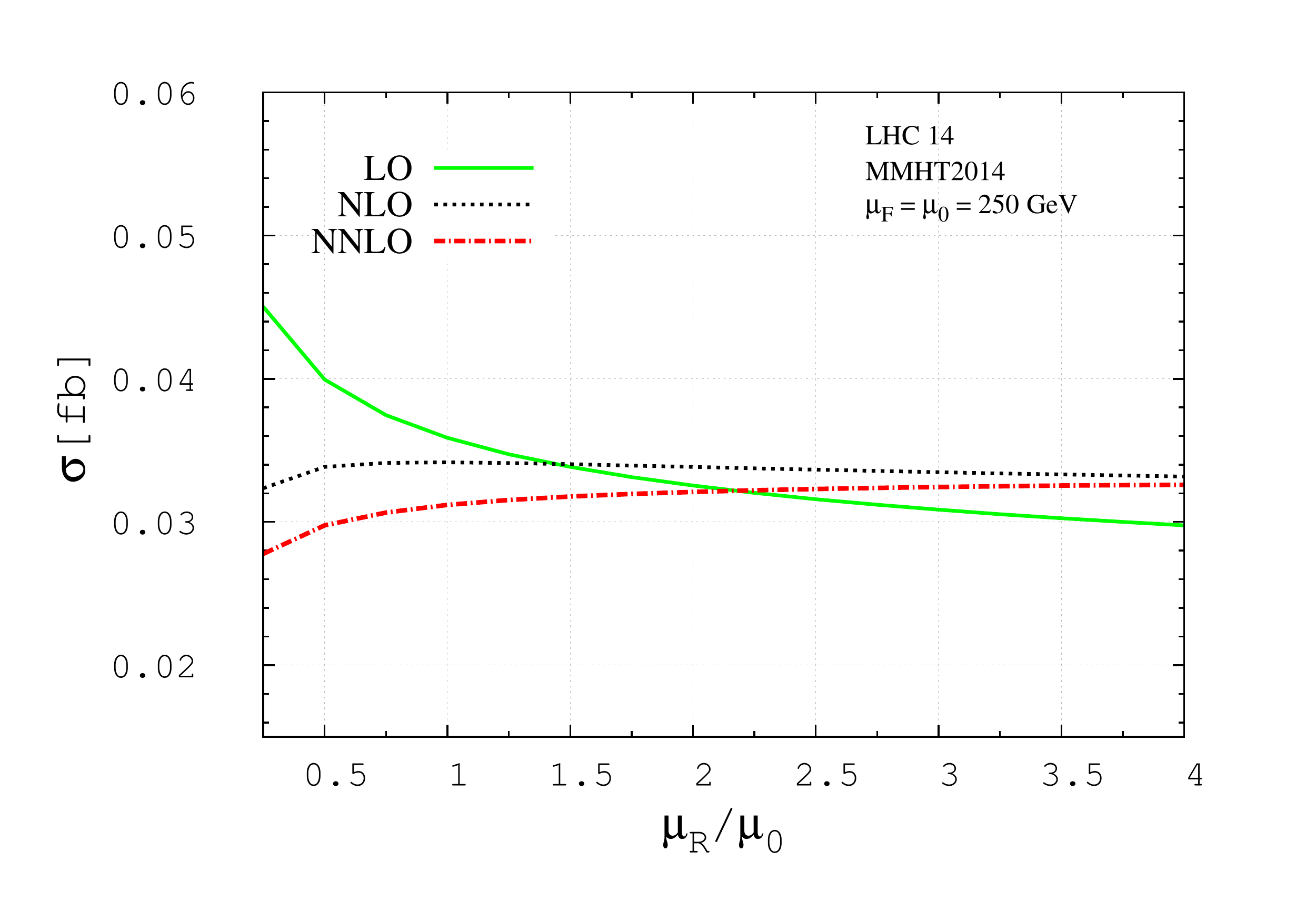}
  \label{fig:sub1}
 \end{subfigure}%
\begin{subfigure}{.5\textwidth}
 \centering
 \includegraphics[width=.75\linewidth, angle=270]{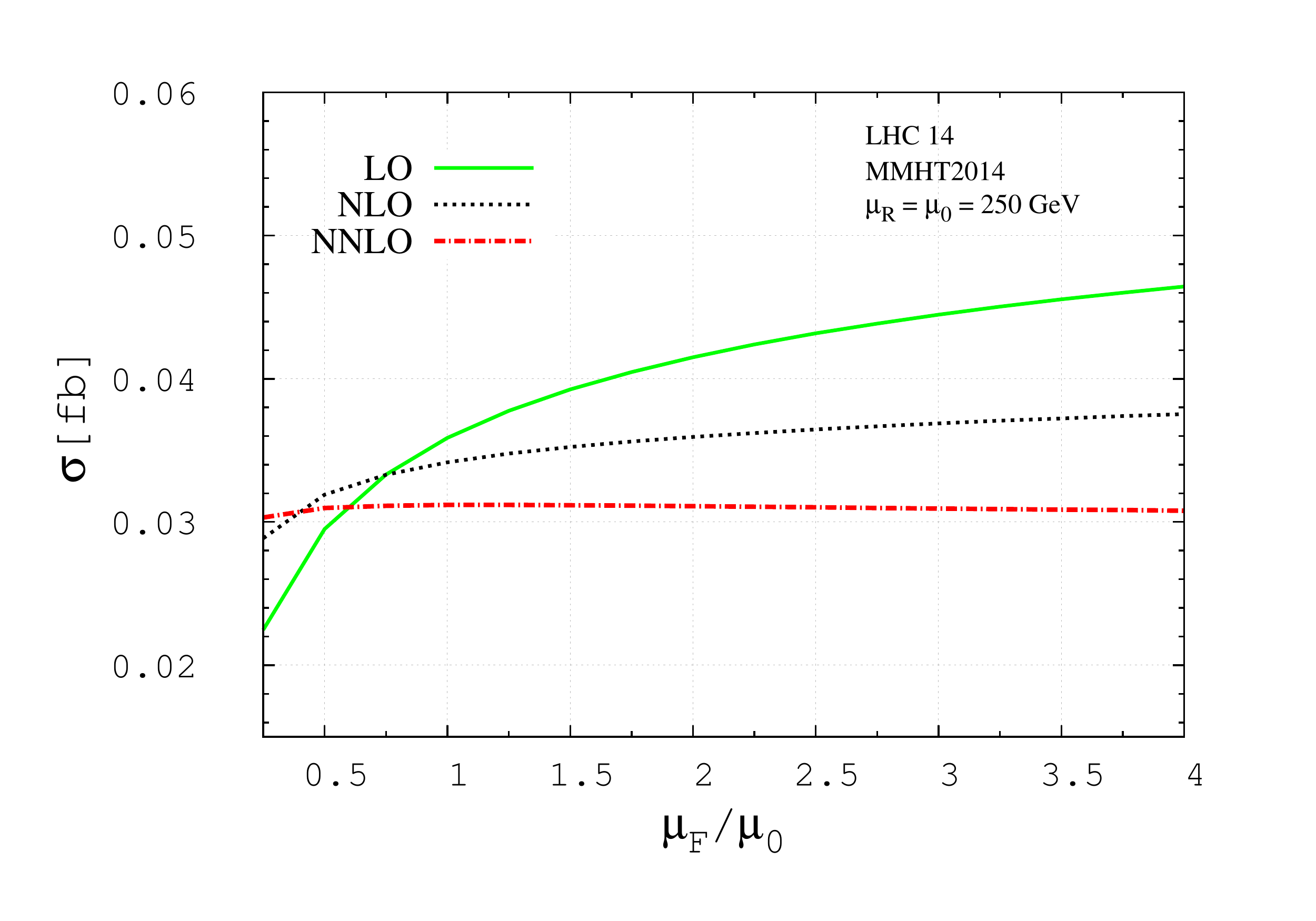}
\label{fig:sub2}
\end{subfigure}
\caption{
\scriptsize{ The total cross section for di-Higgs production in $b\bar{b}$ annihilation at various order in $a_s$ as a function of $\left(\mu_R/\mu_0\right)$ on left panel with $\mu_F = \mu_0$ and as a function of $\left(\mu_F/\mu_0\right)$ on right panel with $\mu_R = \mu_0$  with central scale $\mu_0 = 2 m_h$ and $\sqrt{s} =$ 14 TeV.}}
\label{murmufvar}
\end{figure}

\begin{figure}
\centering

\includegraphics[width=.45\linewidth, angle=270]{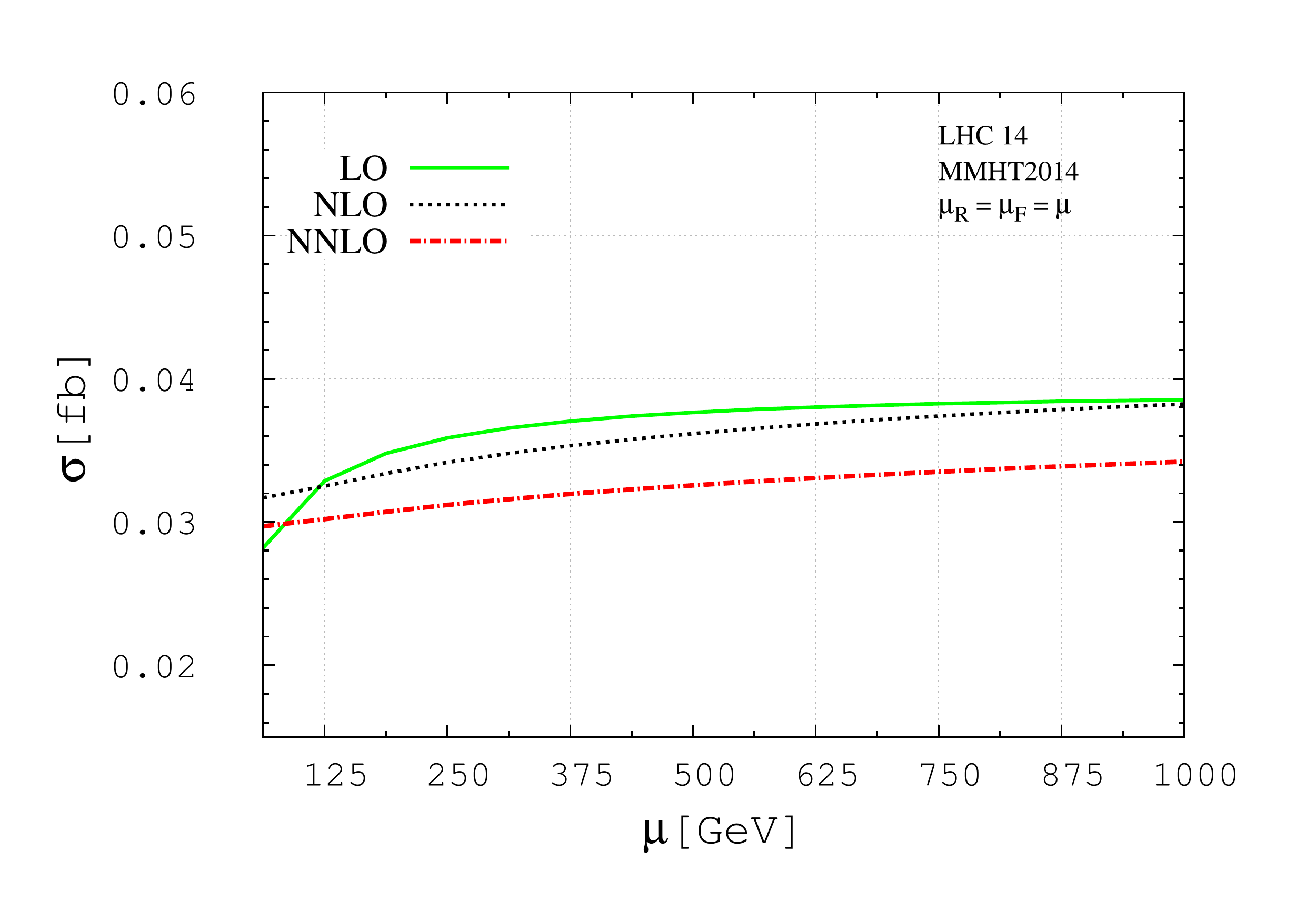} 
\caption{
\scriptsize{ The total cross section for di-Higgs production in $b\bar{b}$ annihilation at various order in $a_s$ as a function of the mass scale $\mu$ with $\left(\mu_F = \mu_R = \mu \right)$ for $\sqrt{s} =$ 14 TeV.}}
\label{muvar}
\end{figure}
In Fig. \ref{murmufvar}, we show the variation of our fixed order predictions with respect to $\mu_R$ (on the left panel) and $\mu_F$ (on the right panel) for a particular choice of central scale $\mu_0 = 250$ GeV. 
We can see that except for the small $\mu_R$ and $\mu_F$ region, which is in the region below $\mu_R = m_h$, there is an overall reduction of the scale dependency with increasing order of perturbation theory.  
We observe that both NLO and NNLO results attain a much faster stability 
against the variation of the scales than the  LO cross-section. 
At the leading order, there are no $\mu_R$ or $\mu_F$ scale dependent logarithms 
that can compensate
those coming from the Yukawa coupling and parton distribution functions,
and hence LO has large scale dependency. 
However, the inclusion of higher order terms that contain logarithms of these scales
provide partial cancellation at every order in perturbation theory. 
Hence the inclusion of NLO and NNLO pieces reduces the dependency on the scales considerably.  
In Fig. \ref{muvar}, we have set $\mu_R = \mu_F$ and varied the cross-section with 
respect to a single scale $\mu$. It can be observed that LO attains stability much 
faster compared to the case when $\mu_R$ is not equal to $\mu_F$. 
This can be comprehended from Fig.~\ref{murmufvar}, 
where the LO contribution behaves exactly in an opposite way  
with respect to the variation of both the mass scales. 
So the stability in the leading order seen in Fig.~\ref{muvar} 
attributes to the fact that there is a significant cancellation 
happening between the $\mu_R$ and $\mu_F$ scale variations of the cross section. 
We also show the 7-point scale variation for the central scale at $m_h=125$ GeV in Table~\ref{tab:table1}. This  variation spans the entire region from $\mu_R, \mu_F = m_h/2$ to $\mu_R, \mu_F = 2 m_h$ and 
hence captures the uncertainty in this region. The 7-point scale variation for a different value of central scale is also shown in Table~\ref{tab:table2}. Table~\ref{tab:table3} contains the $\%$ uncertainty 
from the scale variation at two different central scales. 
It can be seen that the leading order cross-section has a huge scale uncertainty 
which implies the unreliability of the result. 
But the scale dependency starts to reduce when we include the higher order corrections. 

\begin{table}[h!]
  \begin{center}
    \label{tab:table1}
    \begin{tabular}{|P{2.5cm}||P{2.5cm}|P{2.5cm}||P{2.5cm}|} 
      \hline
      $\big(\frac{\mu_R}{\kappa m_h},\frac{\mu_F}{\kappa m_h}\big)$& LO[fb]$\times 10^{-1}$ & NLO[fb]$\times 10^{-1}$ & NNLO[fb]$\times 10^{-1}$\\
     \hline \hline 
        
        (2,2)  & 0.3587& 0.3416& 0.3119\\ 
     
       \cline{1-4}
       (2,1)& 0.2951& 0.3191 & 0.3098 \\
       \cline{1-4}
        (1,2)& 0.3994& 0.3384 & 0.2976\\
       \cline{1-4}
       (1,1)& 0.3286& 0.3250 & 0.3020\\
       \cline{1-4}
      
        (1,1/2)& 0.2502 & 0.3032 & 0.3031\\
       \cline{1-4}
       (1/2,1)&0.3704 & 0.3246 & 0.2879\\
       \cline{1-4}
      (1/2,1/2)& 0.2821 & 0.3169 & 0.2970\\
       \hline 
    \end{tabular}
     \caption{7-point scale variation for central scale at $m_h = 125$GeV,\hspace{2mm}$\kappa=1$}
     \label{tab:table1}
  \end{center}
\end{table}

\begin{table}[h!]
  \begin{center}
    \label{tab:table2}
    \begin{tabular}{|P{2.5cm}||P{2.5cm}|P{2.5cm}||P{2.5cm}|} 
      \hline
    $\big(\frac{\mu_R}{\kappa m_h},\frac{\mu_F}{\kappa m_h}\big)$ & LO[fb]$\times10^{-1}$ & NLO[fb]$\times10^{-1}$ & NNLO[fb]$\times10^{-1}$\\
     \hline \hline 
        
        (2,2)  & 0.3765 & 0.3617 &0.3256 \\ 
     
       \cline{1-4}
       (2,1)& 0.3254 & 0.3384 &0.3210 \\
       \cline{1-4}
        (1,2)& 0.4150 & 0.3594&0.3110 \\
       \cline{1-4}
       (1,1)& 0.3587 & 0.3416 &0.3119 \\
       \cline{1-4}
      
        (1,1/2)& 0.2951 & 0.3191 & 0.3098\\
       \cline{1-4}
       (1/2,1)&0.3994 &0.3384 & 0.2976\\
       \cline{1-4}
      (1/2,1/2)& 0.3286 & 0.3250 & 0.3020\\
       \hline 
    \end{tabular}
     \caption{7-point scale variation for central scale at $m_h = 125$GeV,\hspace{2mm} $\kappa=2$}
      \label{tab:table2}
  \end{center}
\end{table}

\begin{table}[!htb]
 \begin{center}
    \label{tab:table3}
    \begin{tabular}{ | m{2.5cm}||m{2.5cm}|m{2.5cm}||m{2.6cm} | }
 \hline
    Central Scale(GeV) & LO[fb]$\times10^{-1}$ & NLO[fb]$\times10^{-1}$ & NNLO[fb]$\times10^{-1}$\\    
\hline\hline  
        
        125  & $0.3286 ^{+21.546\%}_{-23.859\%}$ & $0.3250^{+5.108\%}_{-6.708\%}$ &$0.3020^{+3.278\%}_{-4.669\%}$ \\ [10pt]
     
      \hline 
       250 & $0.3587^{+15.696\%}_{-17.731\%}$ & $0.3416^{+5.210\%}_{-6.587\%}$ & $0.3119^{+4.392\%}_{-4.585\%}$\\[10pt]
       \hline
\end{tabular}
 \caption{$\%$ scale uncertainty at LO, NLO and NNLO}
 \label{tab:table3}
 \end{center}
\end{table}


\section{Conclusion}
\label{conclusion}
The extraction of the trilinear coupling of the Higgs bosons provides the valuable information on the 
shape of the Higgs potential.  One of the most important observables sensitive to this coupling
is the production of a pair of Higgs bosons at the LHC.  Among various partonic channels that contribute
to this process, gluon fusion is the dominant one and is well studied both in effective theory
as well as in the full theory.  In the effective theory, top quarks are integrated out.
As the precision at the hadron collider improves, it is important to incorporate other sub-dominant channels
to the production mechanism.  In this paper, we have considered one such channel, namely the production of
a pair of Higgs bosons in the bottom quark annihilation which is sensitive to the trilinear coupling.  
Both LO and NLO QCD contributions exist in the literature and hence we  
have computed the NNLO QCD corrections to the inclusive cross section.   
There are two classes of diagrams that contribute at two-loops as well as at real emission
sub processes. The vertex type of virtual diagrams 
that belongs to class-A is already known up to three loops and hence we evaluate only the Class-B diagrams
up to two loops in perturabative QCD for our NNLO predictions. 
Our results are expressed in terms of classical polylogarithms of weight up to 4.  
We observe that the infrared poles of the amplitudes are in agreement with the predictions by Catani. 
We have studied the numerical stability of the coefficient $\mathcal{C}_2^{(2),\rm fin}$ over the 
range of $x$ and $\cos(\theta)$ required for further phenomenological studies.

Using these results at hand, we have computed full NNLO corrections from class-A diagrams and
SV contributions at NNLO level from class-B diagrams to the inclusive cross section.
We have shown how the NNLO results that are already available for the single Higgs boson production
can be used to obtain the full NNLO contributions for the production of pair of Higgs bosons from
class-A diagrams.  Exploiting the universal structure of the soft contributions and the two loop amplitudes
computed in this paper, we derive the SV contributions from class-B diagrams up to NNLO level. 
Our detailed study on the numerical impact of these results at the LHC energy
demonstrates that the inclusion of higher order terms reduces the
renormalization and factorization scale uncertainties making the predictions more reliable.

\section*{Acknowledgements}
We sincerely thank the referee for the valuable comments and suggestions. 
We would like to thank Neelima Agarwal, Sophia Borowka, Goutam Das, Thomas Gehrmann, M. C. Kumar and Anurag Tripathi 
for several useful discussions. We thank Johannes Bl\"umlein for carefully reading the manuscript and providing us with his valuable comments. Ajjath AH and Pooja Mukherjee would like to thank their parents, siblings and friends for the continuous love and support.

\bibliography{main}{}

\providecommand{\href}[2]{#2}\begingroup\raggedright\begin{thebibliography}{10}

\bibitem{Chatrchyan:2012xdj}
{\scshape CMS} collaboration, S.~Chatrchyan et~al., \emph{{Observation of a new
  boson at a mass of 125 GeV with the CMS experiment at the LHC}},
  \href{https://doi.org/10.1016/j.physletb.2012.08.021}{\emph{Phys. Lett.}
  {\bfseries B716} (2012) 30--61},
  [\href{https://arxiv.org/abs/1207.7235}{{\ttfamily 1207.7235}}].

\bibitem{Aad:2012tfa}
{\scshape ATLAS} collaboration, G.~Aad et~al., \emph{{Observation of a new
  particle in the search for the Standard Model Higgs boson with the ATLAS
  detector at the LHC}},
  \href{https://doi.org/10.1016/j.physletb.2012.08.020}{\emph{Phys. Lett.}
  {\bfseries B716} (2012) 1--29},
  [\href{https://arxiv.org/abs/1207.7214}{{\ttfamily 1207.7214}}].

\bibitem{Englert:2014uua}
C.~Englert, A.~Freitas, M.~M. Mühlleitner, T.~Plehn, M.~Rauch, M.~Spira and
  K.~Walz, \emph{{Precision Measurements of Higgs Couplings: Implications for
  New Physics Scales}},
  \href{https://doi.org/10.1088/0954-3899/41/11/113001}{\emph{J. Phys.}
  {\bfseries G41} (2014) 113001},
  [\href{https://arxiv.org/abs/1403.7191}{{\ttfamily 1403.7191}}].

\bibitem{Binoth:2006ym}
T.~Binoth, S.~Karg, N.~Kauer and R.~Ruckl, \emph{{Multi-Higgs boson production
  in the Standard Model and beyond}},
  \href{https://doi.org/10.1103/PhysRevD.74.113008}{\emph{Phys. Rev.}
  {\bfseries D74} (2006) 113008},
  [\href{https://arxiv.org/abs/hep-ph/0608057}{{\ttfamily hep-ph/0608057}}].

\bibitem{ATL-PHYS-PUB-2014-019}
\emph{{Prospects for measuring Higgs pair production in the channel
  $H(\rightarrow\gamma\gamma)H(\rightarrow b\overline{b}) $ using the ATLAS
  detector at the HL-LHC}},  Tech. Rep. ATL-PHYS-PUB-2014-019, CERN, Geneva,
  Oct, 2014.

\bibitem{ATL-PHYS-PUB-2015-046}
\emph{{Higgs Pair Production in the $H(\rightarrow \tau\tau)H(\rightarrow
  b\bar{b})$ channel at the High-Luminosity LHC}},  Tech. Rep.
  ATL-PHYS-PUB-2015-046, CERN, Geneva, Nov, 2015.

\bibitem{CMS-PAS-FTR-15-002}
{\scshape CMS Collaboration} collaboration, \emph{{Higgs pair production at the
  High Luminosity LHC}},  Tech. Rep. CMS-PAS-FTR-15-002, CERN, Geneva, 2015.

\bibitem{CMS-DP-2016-064}
{\scshape CMS Collaboration} collaboration, \emph{{Updates on Projections of
  Physics Reach with the Upgraded CMS Detector for High Luminosity LHC}}, .

\bibitem{Nilles:1983ge}
H.~P. Nilles, \emph{{Supersymmetry, Supergravity and Particle Physics}},
  \href{https://doi.org/10.1016/0370-1573(84)90008-5}{\emph{Phys. Rept.}
  {\bfseries 110} (1984) 1--162}.

\bibitem{Glover:1987nx}
E.~W.~N. Glover and J.~J. van~der Bij, \emph{{HIGGS BOSON PAIR PRODUCTION VIA
  GLUON FUSION}},
  \href{https://doi.org/10.1016/0550-3213(88)90083-1}{\emph{Nucl. Phys.}
  {\bfseries B309} (1988) 282--294}.

\bibitem{Eboli:1987dy}
O.~J.~P. Eboli, G.~C. Marques, S.~F. Novaes and A.~A. Natale, \emph{{TWIN HIGGS
  BOSON PRODUCTION}},
  \href{https://doi.org/10.1016/0370-2693(87)90381-9}{\emph{Phys. Lett.}
  {\bfseries B197} (1987) 269--272}.

\bibitem{Plehn:1996wb}
T.~Plehn, M.~Spira and P.~M. Zerwas, \emph{{Pair production of neutral Higgs
  particles in gluon-gluon collisions}},
  \href{https://doi.org/10.1016/0550-3213(96)00418-X,
  10.1016/S0550-3213(98)00406-4}{\emph{Nucl. Phys.} {\bfseries B479} (1996)
  46--64}, [\href{https://arxiv.org/abs/hep-ph/9603205}{{\ttfamily
  hep-ph/9603205}}].

\bibitem{Dawson:1998py}
S.~Dawson, S.~Dittmaier and M.~Spira, \emph{{Neutral Higgs boson pair
  production at hadron colliders: QCD corrections}},
  \href{https://doi.org/10.1103/PhysRevD.58.115012}{\emph{Phys. Rev.}
  {\bfseries D58} (1998) 115012},
  [\href{https://arxiv.org/abs/hep-ph/9805244}{{\ttfamily hep-ph/9805244}}].

\bibitem{Djouadi:1999gv}
A.~Djouadi, W.~Kilian, M.~{M\"uhlleitner} and P.~M. Zerwas, \emph{{Testing
  Higgs selfcouplings at $e^+e^-$ linear colliders}},
  \href{https://doi.org/10.1007/s100529900082}{\emph{Eur. Phys. J.} {\bfseries
  C10} (1999) 27--43}, [\href{https://arxiv.org/abs/hep-ph/9903229}{{\ttfamily
  hep-ph/9903229}}].

\bibitem{Djouadi:1999rca}
A.~Djouadi, W.~Kilian, M.~{M\"uhlleitner} and P.~M. Zerwas, \emph{{Production
  of neutral Higgs boson pairs at LHC}},
  \href{https://doi.org/10.1007/s100529900083}{\emph{Eur. Phys. J.} {\bfseries
  C10} (1999) 45--49}, [\href{https://arxiv.org/abs/hep-ph/9904287}{{\ttfamily
  hep-ph/9904287}}].

\bibitem{Muhlleitner:2000jj}
M.~M. {M\"uhlleitner}, \emph{{Higgs particles in the standard model and
  supersymmetric theories}}, Ph.D. thesis, Hamburg U., 2000.
\newblock \href{https://arxiv.org/abs/hep-ph/0008127}{{\ttfamily
  hep-ph/0008127}}.

\bibitem{Grigo:2013rya}
J.~Grigo, J.~Hoff, K.~Melnikov and M.~Steinhauser, \emph{{On the Higgs boson
  pair production at the LHC}},
  \href{https://doi.org/10.1016/j.nuclphysb.2013.06.024}{\emph{Nucl. Phys.}
  {\bfseries B875} (2013) 1--17},
  [\href{https://arxiv.org/abs/1305.7340}{{\ttfamily 1305.7340}}].

\bibitem{Frederix:2014hta}
R.~Frederix, S.~Frixione, V.~Hirschi, F.~Maltoni, O.~Mattelaer, P.~Torrielli,
  E.~Vryonidou and M.~Zaro, \emph{{Higgs pair production at the LHC with NLO
  and parton-shower effects}},
  \href{https://doi.org/10.1016/j.physletb.2014.03.026}{\emph{Phys. Lett.}
  {\bfseries B732} (2014) 142--149},
  [\href{https://arxiv.org/abs/1401.7340}{{\ttfamily 1401.7340}}].

\bibitem{Maltoni:2014eza}
F.~Maltoni, E.~Vryonidou and M.~Zaro, \emph{{Top-quark mass effects in double
  and triple Higgs production in gluon-gluon fusion at NLO}},
  \href{https://doi.org/10.1007/JHEP11(2014)079}{\emph{JHEP} {\bfseries 11}
  (2014) 079}, [\href{https://arxiv.org/abs/1408.6542}{{\ttfamily 1408.6542}}].

\bibitem{Degrassi:2016vss}
G.~Degrassi, P.~P. Giardino and R.~Gröber, \emph{{On the two-loop virtual QCD
  corrections to Higgs boson pair production in the Standard Model}},
  \href{https://doi.org/10.1140/epjc/s10052-016-4256-9}{\emph{Eur. Phys. J.}
  {\bfseries C76} (2016) 411},
  [\href{https://arxiv.org/abs/1603.00385}{{\ttfamily 1603.00385}}].

\bibitem{Grober:2017uho}
R.~Grober, A.~Maier and T.~Rauh, \emph{{Reconstruction of top-quark mass
  effects in Higgs pair production and other gluon-fusion processes}},
  \href{https://doi.org/10.1007/JHEP03(2018)020}{\emph{JHEP} {\bfseries 03}
  (2018) 020}, [\href{https://arxiv.org/abs/1709.07799}{{\ttfamily
  1709.07799}}].

\bibitem{Bonciani:2018omm}
R.~Bonciani, G.~Degrassi, P.~P. Giardino and R.~Grober, \emph{{An Analytical
  Method for the NLO QCD Corrections to Double-Higgs Production}},
  \href{https://arxiv.org/abs/1806.11564}{{\ttfamily 1806.11564}}.

\bibitem{Borowka:2016ehy}
S.~Borowka, N.~Greiner, G.~Heinrich, S.~Jones, M.~Kerner, J.~Schlenk,
  U.~Schubert and T.~Zirke, \emph{{Higgs Boson Pair Production in Gluon Fusion
  at Next-to-Leading Order with Full Top-Quark Mass Dependence}},
  \href{https://doi.org/10.1103/PhysRevLett.117.079901,
  10.1103/PhysRevLett.117.012001}{\emph{Phys. Rev. Lett.} {\bfseries 117}
  (2016) 012001}, [\href{https://arxiv.org/abs/1604.06447}{{\ttfamily
  1604.06447}}].

\bibitem{Borowka:2016ypz}
S.~Borowka, N.~Greiner, G.~Heinrich, S.~P. Jones, M.~Kerner, J.~Schlenk and
  T.~Zirke, \emph{{Full top quark mass dependence in Higgs boson pair
  production at NLO}},
  \href{https://doi.org/10.1007/JHEP10(2016)107}{\emph{JHEP} {\bfseries 10}
  (2016) 107}, [\href{https://arxiv.org/abs/1608.04798}{{\ttfamily
  1608.04798}}].

\bibitem{deFlorian:2013jea}
D.~de~Florian and J.~Mazzitelli, \emph{{Higgs Boson Pair Production at
  Next-to-Next-to-Leading Order in QCD}},
  \href{https://doi.org/10.1103/PhysRevLett.111.201801}{\emph{Phys. Rev. Lett.}
  {\bfseries 111} (2013) 201801},
  [\href{https://arxiv.org/abs/1309.6594}{{\ttfamily 1309.6594}}].

\bibitem{deFlorian:2013uza}
D.~de~Florian and J.~Mazzitelli, \emph{{Two-loop virtual corrections to Higgs
  pair production}},
  \href{https://doi.org/10.1016/j.physletb.2013.06.046}{\emph{Phys. Lett.}
  {\bfseries B724} (2013) 306--309},
  [\href{https://arxiv.org/abs/1305.5206}{{\ttfamily 1305.5206}}].

\bibitem{Grigo:2015dia}
J.~Grigo, J.~Hoff and M.~Steinhauser, \emph{{Higgs boson pair production: top
  quark mass effects at NLO and NNLO}},
  \href{https://doi.org/10.1016/j.nuclphysb.2015.09.012}{\emph{Nucl. Phys.}
  {\bfseries B900} (2015) 412--430},
  [\href{https://arxiv.org/abs/1508.00909}{{\ttfamily 1508.00909}}].

\bibitem{Li:2013flc}
Q.~Li, Q.-S. Yan and X.~Zhao, \emph{{Higgs Pair Production: Improved
  Description by Matrix Element Matching}},
  \href{https://doi.org/10.1103/PhysRevD.89.033015}{\emph{Phys. Rev.}
  {\bfseries D89} (2014) 033015},
  [\href{https://arxiv.org/abs/1312.3830}{{\ttfamily 1312.3830}}].

\bibitem{Maierhofer:2013sha}
P.~Maierhöfer and A.~Papaefstathiou, \emph{{Higgs Boson pair production merged
  to one jet}}, \href{https://doi.org/10.1007/JHEP03(2014)126}{\emph{JHEP}
  {\bfseries 03} (2014) 126},
  [\href{https://arxiv.org/abs/1401.0007}{{\ttfamily 1401.0007}}].

\bibitem{Shao:2013bz}
D.~Y. Shao, C.~S. Li, H.~T. Li and J.~Wang, \emph{{Threshold resummation
  effects in Higgs boson pair production at the LHC}},
  \href{https://doi.org/10.1007/JHEP07(2013)169}{\emph{JHEP} {\bfseries 07}
  (2013) 169}, [\href{https://arxiv.org/abs/1301.1245}{{\ttfamily 1301.1245}}].

\bibitem{deFlorian:2015moa}
D.~de~Florian and J.~Mazzitelli, \emph{{Higgs pair production at
  next-to-next-to-leading logarithmic accuracy at the LHC}},
  \href{https://doi.org/10.1007/JHEP09(2015)053}{\emph{JHEP} {\bfseries 09}
  (2015) 053}, [\href{https://arxiv.org/abs/1505.07122}{{\ttfamily
  1505.07122}}].

\bibitem{Grazzini:2018bsd}
M.~Grazzini, G.~Heinrich, S.~Jones, S.~Kallweit, M.~Kerner, J.~M. Lindert and
  J.~Mazzitelli, \emph{{Higgs boson pair production at NNLO with top quark mass
  effects}}, \href{https://doi.org/10.1007/JHEP05(2018)059}{\emph{JHEP}
  {\bfseries 05} (2018) 059},
  [\href{https://arxiv.org/abs/1803.02463}{{\ttfamily 1803.02463}}].

\bibitem{Banerjee:2018lfq}
P.~Banerjee, S.~Borowka, P.~K. Dhani, T.~Gehrmann and V.~Ravindran,
  \emph{{Two-loop massless QCD corrections to the $g+g \rightarrow H+H$
  four-point amplitude}}, {\emph{Submitted to: JHEP} (2018) },
  [\href{https://arxiv.org/abs/1809.05388}{{\ttfamily 1809.05388}}].

\bibitem{Aivazis:1993pi}
M.~A.~G. Aivazis, J.~C. Collins, F.~I. Olness and W.-K. Tung,
  \emph{{Leptoproduction of heavy quarks. 2. A Unified QCD formulation of
  charged and neutral current processes from fixed target to collider
  energies}}, \href{https://doi.org/10.1103/PhysRevD.50.3102}{\emph{Phys. Rev.}
  {\bfseries D50} (1994) 3102--3118},
  [\href{https://arxiv.org/abs/hep-ph/9312319}{{\ttfamily hep-ph/9312319}}].

\bibitem{Collins:1998rz}
J.~C. Collins, \emph{{Hard scattering factorization with heavy quarks: A
  General treatment}},
  \href{https://doi.org/10.1103/PhysRevD.58.094002}{\emph{Phys. Rev.}
  {\bfseries D58} (1998) 094002},
  [\href{https://arxiv.org/abs/hep-ph/9806259}{{\ttfamily hep-ph/9806259}}].

\bibitem{Kramer:2000hn}
M.~Krämer, F.~I. Olness and D.~E. Soper, \emph{{Treatment of heavy quarks in
  deeply inelastic scattering}},
  \href{https://doi.org/10.1103/PhysRevD.62.096007}{\emph{Phys. Rev.}
  {\bfseries D62} (2000) 096007},
  [\href{https://arxiv.org/abs/hep-ph/0003035}{{\ttfamily hep-ph/0003035}}].

\bibitem{Dicus:1988cx}
D.~A. Dicus and S.~Willenbrock, \emph{{Higgs Boson Production from Heavy Quark
  Fusion}}, \href{https://doi.org/10.1103/PhysRevD.39.751}{\emph{Phys. Rev.}
  {\bfseries D39} (1989) 751}.

\bibitem{Dicus:1998hs}
D.~Dicus, T.~Stelzer, Z.~Sullivan and S.~Willenbrock, \emph{{Higgs boson
  production in association with bottom quarks at next-to-leading order}},
  \href{https://doi.org/10.1103/PhysRevD.59.094016}{\emph{Phys. Rev.}
  {\bfseries D59} (1999) 094016},
  [\href{https://arxiv.org/abs/hep-ph/9811492}{{\ttfamily hep-ph/9811492}}].

\bibitem{Maltoni:2003pn}
F.~Maltoni, Z.~Sullivan and S.~Willenbrock, \emph{{Higgs-boson production via
  bottom-quark fusion}},
  \href{https://doi.org/10.1103/PhysRevD.67.093005}{\emph{Phys. Rev.}
  {\bfseries D67} (2003) 093005},
  [\href{https://arxiv.org/abs/hep-ph/0301033}{{\ttfamily hep-ph/0301033}}].

\bibitem{Olness:1987ep}
F.~I. Olness and W.-K. Tung, \emph{{When Is a Heavy Quark Not a Parton? Charged
  Higgs Production and Heavy Quark Mass Effects in the QCD Based Parton
  Model}}, \href{https://doi.org/10.1016/0550-3213(88)90129-0}{\emph{Nucl.
  Phys.} {\bfseries B308} (1988) 813}.

\bibitem{Gunion:1986pe}
J.~F. Gunion, H.~E. Haber, F.~E. Paige, W.-K. Tung and S.~S.~D. Willenbrock,
  \emph{{Neutral and Charged Higgs Detection: Heavy Quark Fusion, Top Quark
  Mass Dependence and Rare Decays}},
  \href{https://doi.org/10.1016/0550-3213(87)90600-6}{\emph{Nucl. Phys.}
  {\bfseries B294} (1987) 621}.

\bibitem{Harlander:2003ai}
R.~V. Harlander and W.~B. Kilgore, \emph{{Higgs boson production in bottom
  quark fusion at next-to-next-to leading order}},
  \href{https://doi.org/10.1103/PhysRevD.68.013001}{\emph{Phys. Rev.}
  {\bfseries D68} (2003) 013001},
  [\href{https://arxiv.org/abs/hep-ph/0304035}{{\ttfamily hep-ph/0304035}}].

\bibitem{Ahmed:2014pka}
T.~Ahmed, M.~Mahakhud, P.~Mathews, N.~Rana and V.~Ravindran, \emph{{Two-loop
  QCD corrections to Higgs $\to b+\overline{b}+g$ amplitude}},
  \href{https://doi.org/10.1007/JHEP08(2014)075}{\emph{JHEP} {\bfseries 08}
  (2014) 075}, [\href{https://arxiv.org/abs/1405.2324}{{\ttfamily 1405.2324}}].

\bibitem{Gehrmann:2014vha}
T.~Gehrmann and D.~Kara, \emph{{The $Hb\bar{b}$ form factor to three loops in
  QCD}}, \href{https://doi.org/10.1007/JHEP09(2014)174}{\emph{JHEP} {\bfseries
  09} (2014) 174}, [\href{https://arxiv.org/abs/1407.8114}{{\ttfamily
  1407.8114}}].

\bibitem{Ahmed:2014cha}
T.~Ahmed, N.~Rana and V.~Ravindran, \emph{{Higgs boson production through $b
  \bar b$ annihilation at threshold in N$^3$LO QCD}},
  \href{https://doi.org/10.1007/JHEP10(2014)139}{\emph{JHEP} {\bfseries 10}
  (2014) 139}, [\href{https://arxiv.org/abs/1408.0787}{{\ttfamily 1408.0787}}].

\bibitem{Ahmed:2014era}
T.~Ahmed, M.~K. Mandal, N.~Rana and V.~Ravindran, \emph{{Higgs Rapidity
  Distribution in $b {\bar b}$ Annihilation at Threshold in N$^{3}$LO QCD}},
  \href{https://doi.org/10.1007/JHEP02(2015)131}{\emph{JHEP} {\bfseries 02}
  (2015) 131}, [\href{https://arxiv.org/abs/1411.5301}{{\ttfamily 1411.5301}}].

\bibitem{Buza:1996wv}
M.~Buza, Y.~Matiounine, J.~Smith and W.~L. van Neerven, \emph{{Charm
  electroproduction viewed in the variable flavor number scheme versus fixed
  order perturbation theory}},
  \href{https://doi.org/10.1007/BF01245820}{\emph{Eur. Phys. J.} {\bfseries C1}
  (1998) 301--320}, [\href{https://arxiv.org/abs/hep-ph/9612398}{{\ttfamily
  hep-ph/9612398}}].

\bibitem{Bierenbaum:2009mv}
I.~Bierenbaum, J.~Blumlein and S.~Klein, \emph{{Mellin Moments of the
  O(alpha**3(s)) Heavy Flavor Contributions to unpolarized Deep-Inelastic
  Scattering at Q**2 >> m**2 and Anomalous Dimensions}},
  \href{https://doi.org/10.1016/j.nuclphysb.2009.06.005}{\emph{Nucl. Phys.}
  {\bfseries B820} (2009) 417--482},
  [\href{https://arxiv.org/abs/0904.3563}{{\ttfamily 0904.3563}}].

\bibitem{Ablinger:2017err}
J.~Ablinger, J.~Blümlein, A.~De~Freitas, A.~Hasselhuhn, C.~Schneider and
  F.~Wißbrock, \emph{{Three Loop Massive Operator Matrix Elements and
  Asymptotic Wilson Coefficients with Two Different Masses}},
  \href{https://doi.org/10.1016/j.nuclphysb.2017.05.017}{\emph{Nucl. Phys.}
  {\bfseries B921} (2017) 585--688},
  [\href{https://arxiv.org/abs/1705.07030}{{\ttfamily 1705.07030}}].

\bibitem{Blumlein:2018jfm}
J.~Blümlein, A.~De~Freitas, C.~Schneider and K.~Schönwald, \emph{{The
  Variable Flavor Number Scheme at Next-to-Leading Order}},
  \href{https://doi.org/10.1016/j.physletb.2018.05.054}{\emph{Phys. Lett.}
  {\bfseries B782} (2018) 362--366},
  [\href{https://arxiv.org/abs/1804.03129}{{\ttfamily 1804.03129}}].

\bibitem{Dawson:2006dm}
S.~Dawson, C.~Kao, Y.~Wang and P.~Williams, \emph{{QCD Corrections to Higgs
  Pair Production in Bottom Quark Fusion}},
  \href{https://doi.org/10.1103/PhysRevD.75.013007}{\emph{Phys. Rev.}
  {\bfseries D75} (2007) 013007},
  [\href{https://arxiv.org/abs/hep-ph/0610284}{{\ttfamily hep-ph/0610284}}].

\bibitem{Dawson:2007wh}
S.~Dawson, C.~Kao and Y.~Wang, \emph{{SUSY QCD Corrections to Higgs Pair
  Production from Bottom Quark Fusion}},
  \href{https://doi.org/10.1103/PhysRevD.77.113005}{\emph{Phys. Rev.}
  {\bfseries D77} (2008) 113005},
  [\href{https://arxiv.org/abs/0710.4331}{{\ttfamily 0710.4331}}].

\bibitem{HongSheng:2005uy}
H.-S. Hou, W.-G. Ma, R.-Y. Zhang, Y.~Jiang, L.~Han and L.-R. Xing, \emph{{Pair
  production of charged Higgs bosons from bottom-quark fusion}},
  \href{https://doi.org/10.1103/PhysRevD.71.075014}{\emph{Phys. Rev.}
  {\bfseries D71} (2005) 075014},
  [\href{https://arxiv.org/abs/hep-ph/0502214}{{\ttfamily hep-ph/0502214}}].

\bibitem{Liu:2004pv}
J.-J. Liu, W.-G. Ma, G.~Li, R.-Y. Zhang and H.-S. Hou, \emph{{Higgs boson pair
  production in the little Higgs model at hadron collider}},
  \href{https://doi.org/10.1103/PhysRevD.70.015001}{\emph{Phys. Rev.}
  {\bfseries D70} (2004) 015001},
  [\href{https://arxiv.org/abs/hep-ph/0404171}{{\ttfamily hep-ph/0404171}}].

\bibitem{Catani:1998bh}
S.~Catani, \emph{{The Singular behavior of QCD amplitudes at two loop order}},
  \href{https://doi.org/10.1016/S0370-2693(98)00332-3}{\emph{Phys. Lett.}
  {\bfseries B427} (1998) 161--171},
  [\href{https://arxiv.org/abs/hep-ph/9802439}{{\ttfamily hep-ph/9802439}}].

\bibitem{Ravindran:2006cg}
V.~Ravindran, \emph{{Higher-order threshold effects to inclusive processes in
  QCD}}, \href{https://doi.org/10.1016/j.nuclphysb.2006.06.025}{\emph{Nucl.
  Phys.} {\bfseries B752} (2006) 173--196},
  [\href{https://arxiv.org/abs/hep-ph/0603041}{{\ttfamily hep-ph/0603041}}].

\bibitem{Gehrmann:2013cxs}
T.~Gehrmann, L.~Tancredi and E.~Weihs, \emph{{Two-loop master integrals for $q
  \bar{q} \to VV$: the planar topologies}},
  \href{https://doi.org/10.1007/JHEP08(2013)070}{\emph{JHEP} {\bfseries 08}
  (2013) 070}, [\href{https://arxiv.org/abs/1306.6344}{{\ttfamily 1306.6344}}].

\bibitem{Nogueira:1991ex}
P.~Nogueira, \emph{{Automatic Feynman graph generation}},
  \href{https://doi.org/10.1006/jcph.1993.1074}{\emph{J. Comput. Phys.}
  {\bfseries 105} (1993) 279--289}.

\bibitem{Vermaseren:2000nd}
J.~A.~M. Vermaseren, \emph{{New features of FORM}},
  \href{https://arxiv.org/abs/math-ph/0010025}{{\ttfamily math-ph/0010025}}.

\bibitem{vonManteuffel:2012np}
A.~von Manteuffel and C.~Studerus, \emph{{Reduze 2 - Distributed Feynman
  Integral Reduction}},  \href{https://arxiv.org/abs/1201.4330}{{\ttfamily
  1201.4330}}.

\bibitem{Tkachov:1981wb}
F.~V. Tkachov, \emph{{A Theorem on Analytical Calculability of Four Loop
  Renormalization Group Functions}},
  \href{https://doi.org/10.1016/0370-2693(81)90288-4}{\emph{Phys. Lett.}
  {\bfseries 100B} (1981) 65--68}.

\bibitem{Chetyrkin:1981qh}
K.~G. Chetyrkin and F.~V. Tkachov, \emph{{Integration by Parts: The Algorithm
  to Calculate beta Functions in 4 Loops}},
  \href{https://doi.org/10.1016/0550-3213(81)90199-1}{\emph{Nucl. Phys.}
  {\bfseries B192} (1981) 159--204}.

\bibitem{Gehrmann:1999as}
T.~Gehrmann and E.~Remiddi, \emph{{Differential equations for two loop four
  point functions}},
  \href{https://doi.org/10.1016/S0550-3213(00)00223-6}{\emph{Nucl. Phys.}
  {\bfseries B580} (2000) 485--518},
  [\href{https://arxiv.org/abs/hep-ph/9912329}{{\ttfamily hep-ph/9912329}}].

\bibitem{Lee:2013mka}
R.~N. Lee, \emph{{LiteRed 1.4: a powerful tool for reduction of multiloop
  integrals}}, \href{https://doi.org/10.1088/1742-6596/523/1/012059}{\emph{J.
  Phys. Conf. Ser.} {\bfseries 523} (2014) 012059},
  [\href{https://arxiv.org/abs/1310.1145}{{\ttfamily 1310.1145}}].

\bibitem{Gehrmann:2014bfa}
T.~Gehrmann, A.~von Manteuffel, L.~Tancredi and E.~Weihs, \emph{{The two-loop
  master integrals for $q\overline{q} \to VV$}},
  \href{https://doi.org/10.1007/JHEP06(2014)032}{\emph{JHEP} {\bfseries 06}
  (2014) 032}, [\href{https://arxiv.org/abs/1404.4853}{{\ttfamily 1404.4853}}].

\bibitem{Gross:1973id}
D.~J. Gross and F.~Wilczek, \emph{{Ultraviolet Behavior of Nonabelian Gauge
  Theories}}, \href{https://doi.org/10.1103/PhysRevLett.30.1343}{\emph{Phys.
  Rev. Lett.} {\bfseries 30} (1973) 1343--1346}.

\bibitem{Politzer:1973fx}
H.~D. Politzer, \emph{{Reliable Perturbative Results for Strong
  Interactions?}},
  \href{https://doi.org/10.1103/PhysRevLett.30.1346}{\emph{Phys. Rev. Lett.}
  {\bfseries 30} (1973) 1346--1349}.

\bibitem{Caswell:1974gg}
W.~E. Caswell, \emph{{Asymptotic Behavior of Nonabelian Gauge Theories to Two
  Loop Order}}, \href{https://doi.org/10.1103/PhysRevLett.33.244}{\emph{Phys.
  Rev. Lett.} {\bfseries 33} (1974) 244}.

\bibitem{Jones:1974mm}
D.~R.~T. Jones, \emph{{Two Loop Diagrams in Yang-Mills Theory}},
  \href{https://doi.org/10.1016/0550-3213(74)90093-5}{\emph{Nucl. Phys.}
  {\bfseries B75} (1974) 531}.

\bibitem{Egorian:1978zx}
E.~Egorian and O.~V. Tarasov, \emph{{Two Loop Renormalization of the {QCD} in
  an Arbitrary Gauge}}, {\emph{Teor. Mat. Fiz.} {\bfseries 41} (1979) 26--32}.

\bibitem{Kinoshita:1962ur}
T.~Kinoshita, \emph{{Mass singularities of Feynman amplitudes}},
  \href{https://doi.org/10.1063/1.1724268}{\emph{J. Math. Phys.} {\bfseries 3}
  (1962) 650--677}.

\bibitem{Lee:1964is}
T.~D. Lee and M.~Nauenberg, \emph{{Degenerate Systems and Mass Singularities}},
  \href{https://doi.org/10.1103/PhysRev.133.B1549}{\emph{Phys. Rev.} {\bfseries
  133} (1964) B1549--B1562}.

\bibitem{Collins:1985ue}
J.~C. Collins, D.~E. Soper and G.~F. Sterman, \emph{{Factorization for Short
  Distance Hadron - Hadron Scattering}},
  \href{https://doi.org/10.1016/0550-3213(85)90565-6}{\emph{Nucl. Phys.}
  {\bfseries B261} (1985) 104--142}.

\bibitem{Kidonakis:1998nf}
N.~Kidonakis, G.~Oderda and G.~F. Sterman, \emph{{Evolution of color exchange
  in QCD hard scattering}},
  \href{https://doi.org/10.1016/S0550-3213(98)00441-6}{\emph{Nucl. Phys.}
  {\bfseries B531} (1998) 365--402},
  [\href{https://arxiv.org/abs/hep-ph/9803241}{{\ttfamily hep-ph/9803241}}].

\bibitem{Sen:1982bt}
A.~Sen, \emph{{Asymptotic Behavior of the Wide Angle On-Shell Quark Scattering
  Amplitudes in Nonabelian Gauge Theories}},
  \href{https://doi.org/10.1103/PhysRevD.28.860}{\emph{Phys. Rev.} {\bfseries
  D28} (1983) 860}.

\bibitem{Garland:2001tf}
L.~W. Garland, T.~Gehrmann, E.~W.~N. Glover, A.~Koukoutsakis and E.~Remiddi,
  \emph{{The Two loop QCD matrix element for e+ e- ---> 3 jets}},
  \href{https://doi.org/10.1016/S0550-3213(02)00057-3}{\emph{Nucl. Phys.}
  {\bfseries B627} (2002) 107--188},
  [\href{https://arxiv.org/abs/hep-ph/0112081}{{\ttfamily hep-ph/0112081}}].

\bibitem{Anastasiou:2001sv}
C.~Anastasiou, E.~W.~N. Glover, C.~Oleari and M.~E. Tejeda-Yeomans, \emph{{Two
  loop QCD corrections to massless quark gluon scattering}},
  \href{https://doi.org/10.1016/S0550-3213(01)00195-X}{\emph{Nucl. Phys.}
  {\bfseries B605} (2001) 486--516},
  [\href{https://arxiv.org/abs/hep-ph/0101304}{{\ttfamily hep-ph/0101304}}].

\bibitem{Glover:2001af}
E.~W.~N. Glover, C.~Oleari and M.~E. Tejeda-Yeomans, \emph{{Two loop QCD
  corrections to gluon-gluon scattering}},
  \href{https://doi.org/10.1016/S0550-3213(01)00210-3}{\emph{Nucl. Phys.}
  {\bfseries B605} (2001) 467--485},
  [\href{https://arxiv.org/abs/hep-ph/0102201}{{\ttfamily hep-ph/0102201}}].

\bibitem{Bern:2002tk}
Z.~Bern, A.~De~Freitas and L.~J. Dixon, \emph{{Two loop helicity amplitudes for
  gluon-gluon scattering in QCD and supersymmetric Yang-Mills theory}},
  \href{https://doi.org/10.1088/1126-6708/2002/03/018}{\emph{JHEP} {\bfseries
  03} (2002) 018}, [\href{https://arxiv.org/abs/hep-ph/0201161}{{\ttfamily
  hep-ph/0201161}}].

\bibitem{Bern:2004cz}
Z.~Bern, L.~J. Dixon and D.~A. Kosower, \emph{{Two-loop g ---> gg splitting
  amplitudes in QCD}},
  \href{https://doi.org/10.1088/1126-6708/2004/08/012}{\emph{JHEP} {\bfseries
  08} (2004) 012}, [\href{https://arxiv.org/abs/hep-ph/0404293}{{\ttfamily
  hep-ph/0404293}}].

\bibitem{Sterman:2002qn}
G.~F. Sterman and M.~E. Tejeda-Yeomans, \emph{{Multiloop amplitudes and
  resummation}},
  \href{https://doi.org/10.1016/S0370-2693(02)03100-3}{\emph{Phys. Lett.}
  {\bfseries B552} (2003) 48--56},
  [\href{https://arxiv.org/abs/hep-ph/0210130}{{\ttfamily hep-ph/0210130}}].

\bibitem{Aybat:2006wq}
S.~M. Aybat, L.~J. Dixon and G.~F. Sterman, \emph{{The Two-loop anomalous
  dimension matrix for soft gluon exchange}},
  \href{https://doi.org/10.1103/PhysRevLett.97.072001}{\emph{Phys. Rev. Lett.}
  {\bfseries 97} (2006) 072001},
  [\href{https://arxiv.org/abs/hep-ph/0606254}{{\ttfamily hep-ph/0606254}}].

\bibitem{Aybat:2006mz}
S.~M. Aybat, L.~J. Dixon and G.~F. Sterman, \emph{{The Two-loop soft anomalous
  dimension matrix and resummation at next-to-next-to leading pole}},
  \href{https://doi.org/10.1103/PhysRevD.74.074004}{\emph{Phys. Rev.}
  {\bfseries D74} (2006) 074004},
  [\href{https://arxiv.org/abs/hep-ph/0607309}{{\ttfamily hep-ph/0607309}}].

\bibitem{Becher:2009cu}
T.~Becher and M.~Neubert, \emph{{Infrared singularities of scattering
  amplitudes in perturbative QCD}},
  \href{https://doi.org/10.1103/PhysRevLett.102.162001,
  10.1103/PhysRevLett.111.199905}{\emph{Phys. Rev. Lett.} {\bfseries 102}
  (2009) 162001}, [\href{https://arxiv.org/abs/0901.0722}{{\ttfamily
  0901.0722}}].

\bibitem{Gardi:2009qi}
E.~Gardi and L.~Magnea, \emph{{Factorization constraints for soft anomalous
  dimensions in QCD scattering amplitudes}},
  \href{https://doi.org/10.1088/1126-6708/2009/03/079}{\emph{JHEP} {\bfseries
  03} (2009) 079}, [\href{https://arxiv.org/abs/0901.1091}{{\ttfamily
  0901.1091}}].

\bibitem{Harris:2001sx}
B.~W. Harris and J.~F. Owens, \emph{{The Two cutoff phase space slicing
  method}}, \href{https://doi.org/10.1103/PhysRevD.65.094032}{\emph{Phys. Rev.}
  {\bfseries D65} (2002) 094032},
  [\href{https://arxiv.org/abs/hep-ph/0102128}{{\ttfamily hep-ph/0102128}}].

\bibitem{Harland-Lang:2014zoa}
L.~A. Harland-Lang, A.~D. Martin, P.~Motylinski and R.~S. Thorne, \emph{{Parton
  distributions in the LHC era: MMHT 2014 PDFs}},
  \href{https://doi.org/10.1140/epjc/s10052-015-3397-6}{\emph{Eur. Phys. J.}
  {\bfseries C75} (2015) 204},
  [\href{https://arxiv.org/abs/1412.3989}{{\ttfamily 1412.3989}}].

\bibitem{Buckley:2014ana}
A.~Buckley, J.~Ferrando, S.~Lloyd, K.~Nordström, B.~Page, M.~Rüfenacht,
  M.~Schönherr and G.~Watt, \emph{{LHAPDF6: parton density access in the LHC
  precision era}},
  \href{https://doi.org/10.1140/epjc/s10052-015-3318-8}{\emph{Eur. Phys. J.}
  {\bfseries C75} (2015) 132},
  [\href{https://arxiv.org/abs/1412.7420}{{\ttfamily 1412.7420}}].

\end{thebibliography}\endgroup
\bibliographystyle{JHEP}
\end{document}